\documentclass[12pt]{article}

\usepackage[margin=2cm]{geometry}
\usepackage{setspace}
\usepackage{amsmath}
\usepackage{float}
\allowdisplaybreaks
\usepackage{eurosym}
\usepackage{comment}
\usepackage{color}
\usepackage[toc,page]{appendix}
\usepackage{graphicx}
\usepackage{multirow}
\usepackage{multicol}
\usepackage{subfigure,comment}
\usepackage{array}
\usepackage{psfrag}
\usepackage{caption}
\usepackage{subfigure}
\usepackage{enumitem}

\usepackage{natbib}
\providecommand{\keywords}[1]
{
  \small	
  \textbf{\textit{Keywords---}} #1
}

\begin{document}

\title{Utility-Scale Energy Storage in an \\ Imperfectly Competitive Power Sector}
\author{Vilma Virasjoki$^{1}$, Afzal Siddiqui$^{1}$ $^{2}$, Fabricio Oliveira$^{1}$, Ahti Salo$^{1}$ \\
        \small $^{1}$Aalto University, Finland \\
        \small $^{2}$University College London, U.K.; Stockholm University, Sweden \\}
\date{\today}
\maketitle

\abstract{Interest in sustainability has increased the share of variable renewable energy sources (VRES) in power generation. Energy storage systems' potential to mitigate intermittencies from non-dispatchable VRES has enhanced their appeal. However, the impacts of storage vary based on the owner and market conditions. We examine the policy implications of investments in utility-scale battery storage via a bi-level optimization model. The lower level depicts power system operations, modeled as either perfect competition or Cournot oligopoly to allow for the assessment of producer market power. The upper-level investor is either a welfare-maximizer or a profit-maximizing standalone merchant to reflect either welfare enhancement or arbitrage, respectively. We implement a realistic case study for Western Europe based on all possible size-location storage investment combinations. We find that market competition affects investment sizes, locations, and their profitability more than the investor's objectives. A welfare-maximizer under perfect competition invests the most in storage capacity. Consumers typically gain most from storage investments in all cases, exceeding the gains for the investors. Specifically, our results show that storage investments may either not occur or be located differently than at social optimum, if market power is exerted. Thus, policy makers need to anticipate producer market power when setting regulation.
}


\section*{Nomenclature}

\subsection*{Indices and Sets}\label{section:sets}

\noindent
$e \in \mathcal{E}:=\{\text{solar}, \text{wind}\}$: variable renewable energy sources (VRES)  \\
$i \in \mathcal{I}$: all companies \\
$i' \in \mathcal{I'} \subset \mathcal{I}$: power producer companies \\
$j \in \mathcal{J} \subset \mathcal{I}$, where $\{j\} = \mathcal{J}$: storage investor of the upper level (a welfare-maximizing investor or a profit-maximizing standalone merchant) \\
$\ell \in \mathcal{L}$: transmission lines \\
$m \in \mathcal{M}$: seasonal clusters (representative weeks of the model) \\
$n \in \mathcal{N}$: power network nodes\\
$t \in \mathcal{T}$: time periods \\
$u \in \mathcal{U}_{n,i'}$: conventional generation units of producer $i' \in \mathcal{I'}$ at network node $n \in \mathcal{N}$ \\
$y \in \mathcal{Y}$: discrete options for storage investment

\subsection*{Parameters}\label{section:param}

\noindent
$A^e_{m,t,n}$: availability factor for VRES type $e \in \mathcal{E}$ at node $n \in \mathcal{N}$ for time period $t \in \mathcal{T}$ and seasonal cluster $m \in \mathcal{M}$ (--)\\
$B_{n,n^{\prime}}$: element ($n,n^\prime$) of node susceptance matrix, where $n, n^\prime \in \mathcal{N}$ ($1/\Omega$)\\ 
$C^{\text{conv}}_{u}$: generation cost of unit $u \in \mathcal{U}_{n,i'}$ (\euro/MWh) \\
$C^{\text{sto}}$: cost of discharge from storage
(\euro/MWh) \\
$D^{\text{int}}_{m,t,n}$: intercept of linear inverse demand function at node $n \in \mathcal{N}$ in time period $t \in \mathcal{T}$ and seasonal cluster $m \in \mathcal{M}$ (\euro/MWh) \\
$D^{\text{slp}}_{m,t,n}$: slope of linear inverse demand function at node $n \in \mathcal{N}$ in time period $t \in \mathcal{T}$ and seasonal cluster $m \in \mathcal{M}$  (\euro/MWh$^2$)\\
$E^\text{in}_i/E^\text{in}_{i'}/E^\text{in}_j$: storage input efficiency of company $i \in \mathcal{I}$ / producer $i' \in \mathcal{I'}$  / the storage investor $j \in \mathcal{J}$ (--)\\
$E^\text{sto}_i/E^\text{sto}_{i'}/E^\text{sto}_j$ : hourly rate of storage decay of company $i \in \mathcal{I}$ / producer $i' \in \mathcal{I'}$  / the storage investor $j \in \mathcal{J}$ (MW/MWh) \\
$\overline{G}^{\text{conv}}_{n,i',u}$: maximum generation capacity of unit $u \in \mathcal{U}_{n,i'}$ from producer $i' \in \mathcal{I'}$ at node $n \in \mathcal{N}$ (MW) \\
$\overline{G}^{e}_{n,i'}$: maximum generation capacity of producer $i' \in \mathcal{I'}$ for VRES type $e \in \mathcal{E}$ at node $n \in \mathcal{N}$ (MW) \\
$H_{\ell,n}$: element ($\ell, n$) of network transfer matrix, where $\ell \in \mathcal{L}$ and $n \in \mathcal{N}$  ($1/\Omega$)\\
$I$: storage investment cost, amortized for one week (\euro/MWh) \\
$K_{\ell}$: maximum capacity of power line $\ell \in \mathcal{L}$ (MW) \\
$\overline{R}^d_{y}$: available discrete capacities for the storage investment (MWh) \\
$R^{\text{in}}_i/R^{\text{in}}_{i'}/R^{\text{in}}_j$: maximum hourly rate at which storage can be charged by company $i \in \mathcal{I}$ / producer $i' \in \mathcal{I'}$  / the storage investor $j \in \mathcal{J}$ (MW/MWh) \\
$R^{\text{out}}_i/R^{\text{out}}_{i'}/R^{\text{out}}_j$: maximum hourly rate at which storage can be discharged by company $i \in \mathcal{I}$ / producer $i' \in \mathcal{I'}$  / the storage investor $j \in \mathcal{J}$ (MW/MWh) \\
$\overline{R}_{n,i'}$: maximum storage capacity of producer $i' \in \mathcal{I'}$ at node $n \in \mathcal{N}$ (MWh) \\
$\underline{R}_{n,i'}/\underline{R}_{n,j}$: minimum storage capacity factor of producer $i' \in \mathcal{I'}$  / the storage investor $j \in \mathcal{J}$ at node $n \in \mathcal{N}$ (--) 
\\
$R^{\text{up}}_u$: maximum ramp-up rate of generation unit $u \in \mathcal{U}_{n,i'}$ \\
$R^{\text{down}}_u$: maximum ramp-down rate of generation unit $u \in \mathcal{U}_{n,i'}$ \\
$T_t$: duration of period $t \in \mathcal{T}$ (h) \\
$W_m$: weight of seasonal cluster  $m \in \mathcal{M}$ (--)

\subsection*{Primal Variables}\label{section:dv}

\noindent
$g^{\text{conv}}_{m,t,n,i',u}$: generation at node $n \in \mathcal{N}$ by producer $i' \in \mathcal{I'}$ using unit $u \in \mathcal{U}_{n,i'}$ for time period $t \in \mathcal{T}$ and seasonal cluster $m \in \mathcal{M}$ (MWh) \\
$g^{e}_{m,t,n,i'}$: VRES generation of type $e \in \mathcal{E}$ at node $n \in \mathcal{N}$ by producer $i' \in \mathcal{I'}$ for time period $t \in \mathcal{T}$ and seasonal cluster $m \in \mathcal{M}$ (MWh) \\
$q_{m,t,n}$: total sales (demand) at node $n \in \mathcal{N}$ for time period $t \in \mathcal{T}$ and seasonal cluster $m \in \mathcal{M}$ (MWh) \\
$r^{\text{sto}}_{m,t,n,i}/r^{\text{sto}}_{m,t,n,i'}/r^{\text{sto}}_{m,t,n,j}$: (end-of-period) stored energy at node $n \in \mathcal{N}$ for time period $t \in \mathcal{T}$ and seasonal cluster $m \in \mathcal{M}$ by company $i \in \mathcal{I}$ / producer $i' \in \mathcal{I'}$  / the storage investor $j \in \mathcal{J}$ (MWh) \\
$r^{\text{in}}_{m,t,n,i}/r^{\text{in}}_{m,t,n,i'}/r^{\text{in}}_{m,t,n,j}$: energy charged into storage at node $n \in \mathcal{N}$ for time period $t \in \mathcal{T}$ and seasonal cluster $m \in \mathcal{M}$ by company $i \in \mathcal{I}$ / producer $i' \in \mathcal{I'}$  / the storage investor $j \in \mathcal{J}$  (MWh) \\
$r^{\text{out}}_{m,t,n,i}/r^{\text{out}}_{m,t,n,i'}/r^{\text{out}}_{m,t,n,j}$: energy discharged from storage at node $n \in \mathcal{N}$ for time period $t \in \mathcal{T}$ and seasonal cluster $m \in \mathcal{M}$ by company $i \in \mathcal{I}$ / producer $i' \in \mathcal{I'}$  / the storage investor $j \in \mathcal{J}$  (MWh) \\
$v_{m,t,n}$: voltage angle at node $n \in \mathcal{N}$ for time period $t \in \mathcal{T}$ and seasonal cluster $m \in \mathcal{M}$ (rad) \\
$z_{n,j,y}$: binary variable determining the discrete storage investment size at node $n \in \mathcal{N}$, storage investor $j \in \mathcal{J}$, and set of discrete options $y \in \mathcal{Y}$ (-)

\subsection*{Dual Variables}\label{section:dual}

\noindent
$\beta^\text{conv}_{m,t,n,i',u}$: shadow price on generation capacity at node $n \in \mathcal{N}$ for conventional generation unit $u \in \mathcal{U}_{n,i'}$ of producer $i' \in \mathcal{I'}$ for time period $t \in \mathcal{T}$ and seasonal cluster $m \in \mathcal{M}$ (\euro/MWh)\\
$\beta^e_{m,t,n,i'}$: shadow price on VRES capacity at node $n \in \mathcal{N}$ for energy source type $e \in \mathcal{E}$ of producer $i' \in \mathcal{I'}$ for time period $t \in \mathcal{T}$ and seasonal cluster $m \in \mathcal{M}$ (\euro/MWh)\\
$\beta^\text{up}_{m,t,n,i',u}$: shadow price on ramp-up constraint at node $n \in \mathcal{N}$ for conventional generation unit $u \in \mathcal{U}_{n,i'}$ of producer $i' \in \mathcal{I'}$ for time period $t \in \mathcal{T}$ and seasonal cluster $m \in \mathcal{M}$ (\euro/MWh) \\
$\beta^\text{down}_{m,t,n,i',u}$: shadow price on ramp down constraint at node $n \in \mathcal{N}$ for conventional generation unit $u \in \mathcal{U}_{n,i'}$ of producer $i' \in \mathcal{I'}$ for time period $t \in \mathcal{T}$ and seasonal cluster $m \in \mathcal{M}$ (\euro/MWh) \\ 
$\theta_{m,t,n}$: shadow price on energy balance (power price) at node $n \in \mathcal{N}$ for time period $t \in \mathcal{T}$ and seasonal cluster $m \in \mathcal{M}$ (\euro/MWh)\\
$\lambda^\text{bal}_{m,t,n,i}$: shadow price on stored energy balance at node $n \in \mathcal{N}$ of company $i \in \mathcal{I}$ for time period $t \in \mathcal{T}$ and seasonal cluster $m \in \mathcal{M}$ (\euro/MWh)\\
$\lambda^\text{in,p}_{m,t,n,i'}$/$\lambda^\text{in,m}_{m,t,n,j}$: shadow price on maximum storage charging at node $n \in \mathcal{N}$ for time period $t \in \mathcal{T}$ and seasonal cluster $m \in \mathcal{M}$ of producer $i' \in \mathcal{I'}$ / storage investor $j \in \mathcal{J}$ (\euro/MWh)\\
$\lambda^\text{lb,p}_{m,t,n,i'}, \lambda^\text{ub,p}_{m,t,n,i'}$/$\lambda^\text{lb,m}_{m,t,n,j}, \lambda^\text{ub,m}_{m,t,n,j}$: shadow price on energy storage minimum and maximum capacity at node $n \in \mathcal{N}$ for time period $t \in \mathcal{T}$ and seasonal cluster $m \in \mathcal{M}$ of producer $i' \in \mathcal{I'}$ / storage investor $j \in \mathcal{J}$  (\euro/MWh)\\
$\lambda^\text{out,p}_{m,t,n,i'}$/$\lambda^\text{out,m}_{m,t,n,j}$: shadow price on maximum storage discharging  at node $n \in \mathcal{N}$ for time period $t \in \mathcal{T}$ and seasonal cluster $m \in \mathcal{M}$ of producer $i' \in \mathcal{I'}$ / storage investor $j \in \mathcal{J}$ (\euro/MWh)\\
$\overline{\mu}_{m,t,\ell}, \underline{\mu}_{m,t,\ell}$: shadow price on transmission capacity for transmission line $\ell \in \mathcal{L}$ for time period $t \in \mathcal{T}$ in seasonal cluster $m \in \mathcal{M}$ (\euro/MW) 

\section{Introduction}

\subsection{Background and Literature}

Ongoing changes in power markets have increased interest in energy storage. In particular, the emergence of large-scale variable renewable energy sources (VRES) makes it challenging to plan and to operate power systems. Therefore, energy storage has been proposed as a possibility to address the resulting systemic imbalances and uncertainties \citep{EC2009,EC2017_storage}. Energy-storage technologies include mature and large-scale technologies like pumped-hydro storage, which comprises over 95\% of the installed storage capacity worldwide \citep{SANDIA}. However, technological advancements and increased volatility in short-term power supply have made also smaller installations such as electrochemical batteries attractive sources for flexibility \citep{IRENA2017,EC2017_storage}.

Depending on the technology, storage can be used for arbitrage, as ancillary services, for generation and transmission capacity replacement or delay, as well as for the aforementioned VRES support (cf. \cite{Sioshansi2017}). Specifically, storage can be valuable as a competitive, profit-making market asset for power producers or merchants, for instance. However, also regulated entities such as transmission system operators (TSOs) or independent system operators (ISOs) could benefit from owning storage capacity. From an economic viewpoint, storage can increase social welfare by enhancing market efficiency (e.g., \cite{SK2011}). Likewise, from a technical perspective, storage alleviates power network congestion and reduces power plant ramping, even if it is owned by producers and primarily used for profit-making \citep{Virasjoki2016}.

Compared to producer or consumer ownership, \cite{Sioshansi2010} shows how storage owned by a standalone merchant can enhance social welfare. The merchant is a profit-maximizing operator, who does not own power generation capacity but merely participates in power markets with its storage assets. By contrast, consumers overuse and producers underuse their storage capacity; consumers ignore the harm that extensive use of storage may have on producers due to the price-smoothing effect, while producers want to avoid this. Other unintended systemic consequences may also occur if the owners' objectives are not aligned with social optimum or the market design does not internalize all social costs. For instance, storage can increase CO$_2$ emissions if it enables the use of cheap but polluting and inflexible technologies during off-peak hours \citep{LA2014,Virasjoki2016}.

Therefore, the competitiveness and structure of markets affect the situation as well. \citet{Sioshansi2017} discusses how the current market design may not be optimal in supporting the market entry or cost-effectiveness of energy storage. Many markets do not support hybrid participation from multiple value streams or owner roles, which may hamper socially optimal storage investments. Additionally, competition in the market may be imperfect, meaning that some participants use market power for their benefit. For instance, companies that exert market power typically underuse storage assets due to the price-smoothing effect \citep{SK2011}, and strategic use of storage can change the power-flow directions \citep{Virasjoki2016}.

Apart from the impacts of storage that depend on the owners and market settings, it is important to note that the optimal size of a storage investment is endogenous to the markets. \citet{Nasrolahpour2016} take the arbitrage point-of-view under demand and price uncertainty for large-scale pumped-hydro facilities. They find that uncertainty increases the size of the storage investment and its profits. Additionally, an investor that exerts market power invests more in power [MW] capacity and less in energy [MWh] capacity than a welfare-maximizer does.

\citet{Dvorkin2018} and \citet{Xu2017} also consider the optimal location of storage. This is important for capturing the value of storage when the transmission lines are congested, i.e., not only the temporal but also the spatial arbitrage. The profitability of battery storage for a merchant depends on transmission-line investments and whether it enables benefiting from high wind capacity and frequent occurrence of congested lines \citep{Dvorkin2018}. The model by \citet{Xu2017} further accounts for the revenues from providing ancillary services in reserve markets for two types of storage: compressed air and lithium-ion battery from a central-planning perspective. 

In addition to modeling the transmission network, \cite{Gonzalez-Romero2019} also capture the effects of imperfect competition on the lower-level power market. Focusing more on the different operations of the storage investor along with imperfect markets, \cite{Siddiqui2018} formulate a two-period model for energy storage investments. They prove that a merchant invests more in storage capacity than a welfare-maximizer does if the market is perfectly competitive. This way the merchant can profit from a high trading volume. Under Cournot oligopoly, however, the merchant invests less than the welfare-maximizer does to keep price differences high and benefit from temporal arbitrage.

\subsection{Research Objectives and Contribution}

Storage operations and investments may not be socially optimal (i.e., the most economically efficient) due to the conflicting objectives of owners and stakeholders. Furthermore, the state of competition affects whether and where storage would be useful. In a recent paper, \cite{Nasrolahpour2016} ignore market power at the lower level and do not model a network with transmission constraints. Therefore, they do not conduct either welfare or congestion-alleviation analyses. \citet{Xu2017} and \citet{Dvorkin2018} consider also storage locations and transmission line limits but ignore both strategic investors and market power at the lower level. \cite{Siddiqui2018} consider imperfectly competitive markets, but their analytical approach lacks real-world details, such as transmission topology or VRES intermittency. This paper extends the work by \cite{Virasjoki2016}, which studies the market impacts of storage operations as a single-level complementarity problem, to a bi-level model of storage investments and storage ownership.

In this paper, our main objective is to study the optimal storage investment and its welfare effects, assuming different levels of competition and storage investors. Although perfect competition is often assumed in power market models, this may not represent reality, because large power companies tend to own a significant share of national generation capacity. Thus, for the market, we compare (a) perfect competition with (b) Cournot oligopoly, to assess the effects of market power. The storage investor at the upper level can be either (i) an entity that maximizes social welfare or (ii) a merchant who does not own any other assets in the market prior to the possible investment. The main contribution is to develop a comprehensive modeling assessment considering investments with different objectives and at different markets, acknowledging both the transmission constraints and impacts from several seasons.

We reformulate the resulting bi-level optimization models into single-level equivalent models by using a mathematical program with primal and dual constraints (MPPDC) approach \citep{GCFHR2012}. This means that the convex lower-level problems are represented through their primal and dual feasibility constraints and strong-duality condition, along with the  objective and constraints of the upper-level problem. \cite{Baringo2012} use a similar approach to analyze a transmission-line and wind-power investment decisions. \cite{Huppmann2015} apply this method in the lowest (market) level of their tri-level model for nationally strategic transmission capacity investments in Europe. Thereafter, the model can be solved as a mixed integer quadratically constrained quadratic program (MIQCQP) by using discrete storage investment options. We demonstrate the use of this model with a three-node example and note that a welfare-maximizer under perfect competition equals a central-planning paradigm.

Nonetheless, our model cannot be currently solved by most of the commercial and open-source solvers with a realistic large-scale instance. The numerical problems appear to be caused by the solvers' inability to satisfy the quadratic strong-duality condition with a large number of variables and constraints. Therefore, for a case study on Western European power system \citep{Neuhoff2005,GL2010}, we use an iterative approach that requires solving several quadratic programs (QP) instead of an MIQCQP. In effect, this decomposes the investment and operational market decisions. Finally, we account for seasonal variations in VRE generation and demand by using a clustering technique developed by \cite{Reichenberg2019} to create representative weeks on which the investment decisions will be based.

The results for the realistic instance corroborate that especially the market setting (a)-(b), but also the investor type (i)-(ii), affects the optimal investment sizes and locations as well as welfare implications. In contrast to the model in \cite{Siddiqui2018}, the welfare-maximizer invests at least as much as a merchant under perfect competition. Clearly, the optimal investment strategies become less discernible with complexities arising from the generation mix, as well as spatial and longer-term temporal aspects. The optimal new storage locations are in nuclear-dominated Belgium and France. 
However, we note that battery storage for arbitrage purposes only may not be profitable at current price levels. When producers exert market power, storage becomes less attractive as the prices become flatter, albeit higher. Then, a welfare-maximizer typically invests more than a merchant does, but the situation varies depending on the investment price assumption, and the primary investment location shifts to Germany. We show how consumers typically gain most from the investments, exceeding the investor surplus, which may even be negative for a welfare-maximizer.

This paper is structured so that the mathematical MIQCQP model for the storage-investment model is formulated in Section \ref{problemformulation}. Section \ref{numerical_examples} presents illustrative numerical results for a three-node example (Section \ref{3node}) and the case study for Western European test network (Section \ref{fullmodel}). Specifically, Section \ref{problemformulation_alternative} presents the alternative solution approach, in which the QP models are iteratively solved for a discrete set for storage investment options. Section \ref{discussion} concludes with a discussion and a summary.

\section{Model Formulation} \label{problemformulation}

\subsection{Assumptions}

The companies in the power market are denoted by $i \in \mathcal{I}$. Producer companies with conventional or VRES capacity are in the subset $i' \in \mathcal{I'} \subset \mathcal{I}$. The upper-level investor (the welfare-maximizing entity or standalone merchant) is denoted by $j \in \mathcal{J} \subset \mathcal{I}$, where $\{j\} = \mathcal{J}$ (i.e., there is only one upper-level investor within the model) and $j \notin \mathcal{I'}$. Producers can have existing storage assets (mainly pumped-hydro), but they always own other generation capacity as well, whereas the storage-investor operates only storage capacity.

Storage investment is considered from the viewpoint of battery storage, such as lithium-ion. It is available in discrete sizes $\overline{R}^d_{y}$, which are chosen from the set $y \in \mathcal{Y}$ (incl. $\overline{R}^d_{1}$ = 0)  with the binary variable $z_{n,j,y} \in \{ 0,1 \}$. Consequently, the storage investor $j$ acts as a price-taker due to a minor position in a power market dominated by big incumbent companies and large-scale pumped-hydro storage. By contrast, producers $i'$ can either be price-takers or exert market power, depending on the formulation.

VRES production is modeled as deterministic, but we account for multiple seasonal scenarios for demand and VRES production. For this, we use representative weeks, which we obtain by using hierarchical clustering \citep{Reichenberg2019}. Hence, we do not model uncertainty within these weeks (to which e.g., stochastic programming could be used), but we implicitly capture VRE and demand variability on a yearly level. The weights $W_m$ of the clusters $m \in \mathcal{M}$ sum up to one, $\sum_{m \in \mathcal{M}}$ $W_m = 1$ and are based on the number of weeks within the clusters. Furthermore, VRES have priority access to the grid, meaning that all available wind and solar PV (photovoltaic) production is used with zero marginal costs. Finally, power transmission is modeled with DC load-flow linearization.

\subsection{Central Planning (MIQP): Storage Investment and Power Market Operations}

A single-level central-planning problem is formulated as a benchmark prior to the bi-level problems. The central planner decides both on the optimal storage investment and the market operations to maximize social welfare. This is written as follows.
\begin{align}
\max_{\substack{}} &\sum_{m \in \mathcal{M}} \sum_{t \in \mathcal{T}} \sum_{n \in \mathcal{N}} W_{m}   \, \Bigg[  
\bigg( D^{\text{int}}_{m,t,n} q_{m,t,n} - \frac{1}{2} D_{m,t,n}^{\text{slp}} q_{m,t,n}^2 \bigg) 
\nonumber \\
& - \sum_{i'\in \mathcal{I'}} \sum_{u \in \mathcal{U}_{n,i'}}  C_{u}^\text{conv}  g^\text{conv}_{m,t,n,i',u}   -  \sum_{i\in \mathcal{I}} C^{\text{sto}} r^{\text{out}}_{m,t,n,i} \Bigg] - \sum_{n \in \mathcal{N}}  \sum_{y \in \mathcal{Y}}  z_{n,j,y} I\overline{R}^d_{y} \label{swmax_cp}  \\
\text{s.t.} \quad & \sum_{y \in \mathcal{Y}} z_{n,j,y} =1 \,  \forall n, \quad z_{n,j,y} \in \{ 0,1 \} \label{binary_constraint} \\
& \quad  q_{m,t,n} -  \sum_{i'\in \mathcal{I'}} \sum_{\substack{u \in \mathcal{U}_{n,i'}}} g^\text{conv}_{m,t,n,i',u} -  \sum_{i'\in \mathcal{I'}} \sum_{\substack{e \in \mathcal{E}}}  g^e_{m,t,n,i'} -  \sum_{i\in \mathcal{I}} r^\text{out}_{m,t,n,i} \nonumber \\ & + \sum_{i\in \mathcal{I}} r^\text{in}_{m,t,n,i}  - \sum_{\substack{n^\prime \in \mathcal{N}}} T_{t} B_{n,n^\prime} v_{m,t,n^\prime} = 0\; (\theta_{m,t,n}), \, \forall m,t,n \label{prod2} \\
& g^\text{conv}_{m,t,n,i',u} - T_{t}  \, \overline{G}^\text{conv}_{n,i',u} \leq 0 \; (\beta^\text{conv}_{m,t,n,i',u}), \, \forall m,t,n,i',  u \in \mathcal{U}_{n,i'} \label{prod3} \\
& g^\text{conv}_{m,t,n,i',u} - g^\text{conv}_{m,t-1,n,i',u} - T_{t} \, R^{\text{up}}_u \, \overline{G}^\text{conv}_{n,i',u}  \leq 0 \; \nonumber \\
& ({\beta}^{\text{up}}_{m,t,n,i',u}), \,
  \forall m,t,n,i',  u \in \mathcal{U}_{n,i'}  \label{prod4} \\
& g^\text{conv}_{m,t-1,n,i',u} - g^\text{conv}_{m,t,n,i',u} - T_{t} \, R^{\text{down}}_u \, \overline{G}^\text{conv}_{n,i',u}  \leq 0 \; \nonumber \\
& ({\beta}^{\text{down}}_{m,t,n,i',u}), \,
  \forall m,t,n,i',  u \in \mathcal{U}_{n,i'}  \label{prod5} \\
& g^e_{m,t,n,i'} - T_{t} \,  A^e_{m,t,n} \, \overline{G}{}^e_{n,i'} = 0 \; ({\beta}^{e}_{m,t,n,i'}),  \, \forall m,t,n,i', e   \label{prod6} \\
& {r}^{\text{sto}}_{m,t,n,i} -  (1-E^\text{sto}_i)^{T_{t}} \,  r^{\text{sto}}_{m,t-1,n,i}  - E^\text{in}_i \,  r^\text{in}_{m,t,n,i}  + r^\text{out}_{m,t,n,i} = 0
\; \nonumber \\
& ({\lambda}^\text{bal}_{m,t,n,i}) , \, \forall m,t,n,i \label{prod7} \\
& r^\text{in}_{m,t,n,i'} - T_{t} \,  R^\text{in}_{i'} \,  \overline{R}_{n,i'} \leq 0  \; ({\lambda}^\text{in,p}_{m,t,n,i'} ) , \, \forall m,t,n,i'  \label{prod8} \\
& r^\text{out}_{m,t,n,i'} - T_{t} \,  R^\text{out}_{i'} \,  \overline{R}_{n,i'} \leq  0 \; ({\lambda}^\text{out,p}_{m,t,n,i'} ) , \, \forall m,t,n,i'  \label{prod9} \\
&  r^{\text{sto}}_{m,t,n,i'} - \overline{R}_{n,i'} \leq 0 \; ( {\lambda}^\text{ub,p}_{m,t,n,i'} ),  \, \forall m,t,n,i'  \label{prod10} \\
& \underline{R}_{n,i'} \overline{R}_{n,i'} - r^{\text{sto}}_{m,t,n,i'} \leq 0 \; ({\lambda}^\text{lb,p}_{m,t,n,i'}),  \, \forall m,t,n,i'  \label{prod11} \\
& r^\text{in}_{m,t,n,j} - T_{t} \,  R^\text{in}_j \,  \sum_{y \in \mathcal{Y}} z_{n,j,y} \overline{R}^d_{y} \leq 0  \; ({\lambda}^\text{in,m}_{m,t,n,j} ) , \, \forall m,t,n,j  \label{prod12} \\
& r^\text{out}_{m,t,n,j} - T_{t} \,  R^\text{out}_j \,  \sum_{y \in \mathcal{Y}} z_{n,j,y} \overline{R}^d_{y} \leq  0 \; ({\lambda}^\text{out,m}_{m,t,n,j} ) , \, \forall m,t,n,j  \label{prod13} \\
&  r^{\text{sto}}_{m,t,n,j} -\sum_{y \in \mathcal{Y}} z_{n,j,y} \overline{R}^d_{y} \leq 0 \; ( {\lambda}^\text{ub,m}_{m,t,n,j} ),  \, \forall m,t,n,j  \label{prod14} \\
& \underline{R}_{n,j} \sum_{y \in \mathcal{Y}} z_{n,j,y} \overline{R}^d_{y} - r^{\text{sto}}_{m,t,n,j} \leq 0 \; ({\lambda}^\text{lb,m}_{m,t,n,j}),  \, \forall m,t,n,j  \label{prod15} \\
& \sum_{\substack{n \in \mathcal{N}}} T_{t}H_{\ell,n}v_{m,t,n} - T_{t}K_{\ell} \leq 0  \; (\overline{\mu}_{m,t,\ell}), \, \forall m,t,\ell  \label{grid2} \\
& - \sum_{\substack{n \in \mathcal{N}}} T_{t}H_{\ell,n} v_{m,t,n} - T_{t}K_{\ell} \leq 0 \; (\underline{\mu}_{m,t,\ell}), \, \forall m,t,\ell \label{grid3}
\end{align}

and the following variable restrictions:
\begin{align}
g^\text{conv}_{m,t,n,i',u} & \geq 0, \,  \forall m,t,n,i',u \in \mathcal{U}_{n,i'} \label{primalvariables1} \\  
g^e_{m,t,n,i'}  & \geq 0, \,  \forall m,t,n,i',e \label{primalvariables2} \\
r^{\text{sto}}_{m,t,n,i}, r^\text{in}_{m,t,n,i}, r^\text{out}_{m,t,n,i} & \geq 0, \,  \forall m,t,n,i \label{primalvariables3}  \\
q_{m,t,n} \geq 0, \; v_{m,t,n} \, &\text{u.r.s.}, \, \forall m,t,n. \label{primalvariables4} 
\end{align}

The corresponding dual variables are written in brackets after each constraint. Constraint \eqref{prod2} ensures a balance between power demand, generation, storage use, and transmission. Equations \eqref{prod3} - \eqref{prod5} set maximum generation and ramping levels for conventional generation, while \eqref{prod6} defines VRES generation based on its availability. Constraints \eqref{prod7} - \eqref{prod15} account for storage use: both for the existing capacity ($i'$) and for possible new capacity resulting from the investment ($j$). Finally, equations \eqref{grid2} - \eqref{grid3} determine power network operations for the DC load-flow linearization.

\subsection{Decentralized Electricity Industry: The Bi-Level Problem}

In a bi-level setting, the decision of the upper-level decision-maker affects the lower-level problem. Here, the storage investment decision affects the level of installed storage capacity available in power markets, constructing a bi-level model.

\begin{figure}[ht]
  \centering
  \includegraphics[width=11cm]{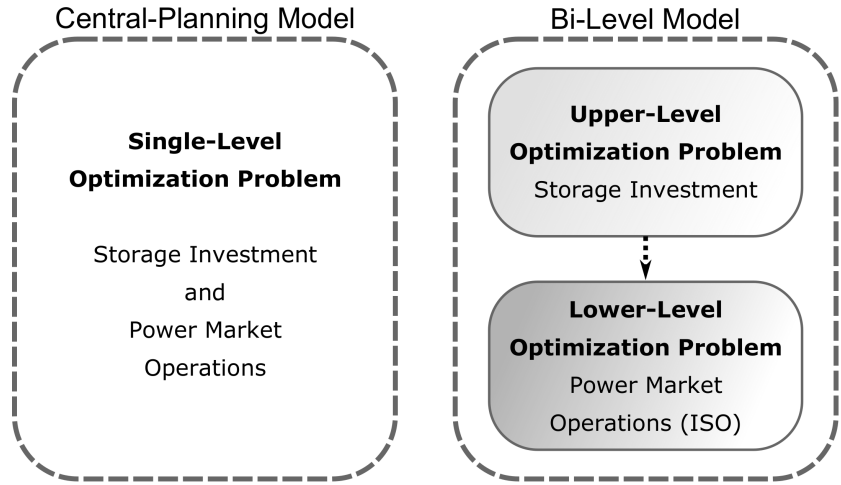}
  \caption{A schematic illustration on the model structures.}
  \label{fig:models}
\end{figure}

\subsubsection{Lower-Level Problem: Power Market Operations}

The power market is modeled as an independent system operator (ISO) who decides on generation levels, storage operations, and grid use. The model is able to reflect both perfect and imperfect competition. Here, the objective function \eqref{iso_co} takes into account the strategic behavior of the producers $i' \in \mathcal{I'} \subset \mathcal{I}$ with \textbf{the extended cost term}\footnote{The extended cost is the second row of \eqref{iso_co} with bold font. Without this term, the formulation corresponds to perfect competition as social welfare maximization.} representing Cournot oligopoly of the markets, i.e., 
\begin{align}
\max_{\substack{}} &\sum_{m \in \mathcal{M}} \sum_{t \in \mathcal{T}} \sum_{n \in \mathcal{N}} W_{m}   \, \Bigg[  
\bigg( D^{\text{int}}_{m,t,n} q_{m,t,n} - \frac{1}{2} D_{m,t,n}^{\text{slp}} q_{m,t,n}^2 \bigg) 
\nonumber \\
& \mathbf{ - \frac{1}{2} D_{m,t,n}^{\textbf{slp}} \sum_{i'\in \mathcal{I'}} \bigg( \sum_{u \in \mathcal{U}_{n,i'}} g^\textbf{conv}_{m,t,n,i',u}  + \sum_{\substack{e \in \mathcal{E}}}  g^e_{m,t,n,i'} + r^\textbf{out}_{m,t,n,i'}  - r^\textbf{in}_{m,t,n,i'} \bigg) ^2 }
\nonumber \\ 
& - \sum_{i'\in \mathcal{I'}} \sum_{u \in \mathcal{U}_{n,i'}}  C_{u}^\text{conv}  g^\text{conv}_{m,t,n,i',u}   -  \sum_{i\in \mathcal{I}} C^{\text{sto}} r^{\text{out}}_{m,t,n,i} \Bigg]
\label{iso_co} \\
\text{s.t.} 
& \quad \text{Equations \eqref{prod2}--\eqref{primalvariables4}}. \nonumber
\end{align}

\subsubsection{Upper-Level Problems: Storage Investment}

The welfare-maximizing storage investor $j$ decides on the optimal storage investment based on the maximization of expected\ social welfare and the investment cost:
\begin{align}
\max_{\substack{}} &\sum_{m \in \mathcal{M}} \sum_{t \in \mathcal{T}} \sum_{n \in \mathcal{N}} W_{m}   \, \Bigg[  
\bigg( D^{\text{int}}_{m,t,n} q_{m,t,n} - \frac{1}{2} D_{m,t,n}^{\text{slp}} q_{m,t,n}^2 \bigg) 
\nonumber \\
& - \sum_{i'\in \mathcal{I}} \sum_{u \in \mathcal{U}_{n,i'}}  C_{u}^\text{conv}  g^\text{conv}_{m,t,n,i',u}   -  \sum_{i\in \mathcal{I}} C^{\text{sto}} r^{\text{out}}_{m,t,n,i} \Bigg] - \sum_{n \in \mathcal{N}} \sum_{y \in \mathcal{Y}} z_{n,j,y} I \overline{R}^d_{y} \label{swmax_upper}  \\
\text{s.t.} & \quad  \text{Equation \eqref{binary_constraint}. \nonumber}
\end{align}

The merchant storage investor $j$ decides on optimal storage investment size and location based on profit maximization:
\begin{align}
\max_{\substack{}} & \sum_{m \in \mathcal{M}} \sum_{t \in \mathcal{T}} \sum_{n \in \mathcal{N}}  \, W_m \bigg[ \frac{\theta_{m,t,n}}{W_m} ( r^{\text{out}}_{m,t,n,j} - r^{\text{in}}_{m,t,n,j}) - C^{\text{sto}} r^{\text{out}}_{m,t,n,j} \bigg] - \sum_{n \in \mathcal{N}} \sum_{y \in \mathcal{Y}} z_{n,j,y} I \overline{R}^d_{y} \label{merchant_upper} \\
\text{s.t.} & \quad  \text{Equation \eqref{binary_constraint}. \nonumber}
\end{align}

Eq. \eqref{merchant_upper} is not, however, convex due to bilinear terms. These comprise products of power prices (lower-level dual variables, $\theta_{m,t,n,}$) and storage operations (lower-level primal variables). By writing the lower-level operational optimization problem of the merchant storage operator (Eq. \eqref{merchant_upper} without the investment cost term, s.t. \eqref{prod12}-\eqref{prod15}) and using linear programming strong duality, we can replace the first term in \eqref{merchant_upper} as
\begin{align}
& \sum_{m \in \mathcal{M}} \sum_{t \in \mathcal{T}} \sum_{n \in \mathcal{N}}   \, W_m \bigg[ \frac{\theta_{m,t,n}}{W_m} ( r^{\text{out}}_{m,t,n,j} - r^{\text{in}}_{m,t,n,j} ) - C^{\text{sto}} r^{\text{out}}_{m,t,n,j} \bigg] 
\nonumber \\
= & \sum_{m \in \mathcal{M}} \sum_{t \in \mathcal{T}} \sum_{n \in \mathcal{N}}  T_{t} \,  R^\text{in}_j \, {\lambda}^\text{in,m}_{m,t,n,j} \sum_{y \in \mathcal{Y}} z_{n,j,y} \overline{R}^d_{y}  + \sum_{m \in \mathcal{M}} \sum_{t \in \mathcal{T}} \sum_{n \in \mathcal{N}}  T_{t} \,  R^\text{out}_j \,  {\lambda}^\text{out,m}_{m,t,n,j} \sum_{y \in \mathcal{Y}} z_{n,j,y} \overline{R}^d_{y} \nonumber \\
& + \sum_{m \in \mathcal{M}} \sum_{t \in \mathcal{T}} \sum_{n \in \mathcal{N}} {\lambda}^\text{ub,m}_{m,t,n,j} \sum_{y \in \mathcal{Y}} z_{n,j,y} \overline{R}^d_{y}  - \sum_{m \in \mathcal{M}} \sum_{t \in \mathcal{T}} \sum_{n \in \mathcal{N}} \underline{R}_{n,j} {\lambda}^\text{lb,m}_{m,t,n,j} \sum_{y \in \mathcal{Y}} z_{n,j,y} \overline{R}^d_{y}.
\label{lpduality}
\end{align}

The right-hand side of Eq. \eqref{lpduality} still includes products of lower-level dual variables and the upper-level storage investment decision. We can reformulate the equation once more by noticing that the same terms appear in the dual objective function of the lower-level ISO (i.e., the primal problem of the ISO being \eqref{iso_co} subject to constraints \eqref{prod2}--\eqref{primalvariables4}). Hence, by using QP strong duality (cf. \cite{Dorn1960,Huppmann2015}) for the convex lower-level problem, we have
\begin{align}
& \sum_{m \in \mathcal{M}} \sum_{t \in \mathcal{T}} \sum_{n \in \mathcal{N}} W_{m}   \, \Bigg[  
\bigg( D^{\text{int}}_{m,t,n} q_{m,t,n} - \frac{1}{2} D_{m,t,n}^{\text{slp}} q_{m,t,n}^2 \bigg) 
 - \frac{1}{2} D_{m,t,n}^{\text{slp}} \sum_{i'\in \mathcal{I'}} \bigg( \sum_{u \in \mathcal{U}_{n,i'}} g^\text{conv}_{m,t,n,i',u} \nonumber \\
& + \sum_{\substack{e \in \mathcal{E}}}  g^e_{m,t,n,i'} + r^\text{out}_{m,t,n,i'}  - r^\text{in}_{m,t,n,i'} \bigg)^2  - \sum_{i'\in \mathcal{I'}} \sum_{u \in \mathcal{U}_{n,i'}}  C_{u}^\text{conv}  g^\text{conv}_{m,t,n,i',u}   -  \sum_{i\in \mathcal{I}} C^{\text{sto}} r^{\text{out}}_{m,t,n,i} \Bigg]
\nonumber \\
= & \sum_{m \in \mathcal{M}} \sum_{t \in \mathcal{T}} \sum_{n \in \mathcal{N}} W_{m}  \, \Bigg[  
 \frac{1}{2} D_{m,t,n}^{\text{slp}} \bigg(  q_{m,t,n}^2 + \sum_{i'\in \mathcal{I'}} \bigg( \sum_{u \in \mathcal{U}_{n,i'}} g^\text{conv}_{m,t,n,i',u}  + \sum_{\substack{e \in \mathcal{E}}}  g^e_{m,t,n,i'} 
 \nonumber \\ 
& + r^\text{out}_{m,t,n,i'}  - r^\text{in}_{m,t,n,i'} \bigg)^2  \bigg)
 \Bigg]
\nonumber \\
& + \sum_{m \in \mathcal{M}} \sum_{t \in \mathcal{T}} \sum_{\ell \in \mathcal{L}} T_{t} K_{\ell}  \big( \underline{\mu}_{m,t,\ell} + \overline{\mu}_{m,t,\ell} \big) - \sum_{m \in \mathcal{M}} \sum_{t \in \mathcal{T}} \sum_{n \in \mathcal{N}} \underline{R}_{n,j} {\lambda}^\text{lb,m}_{m,t,n,j} \sum_{y \in \mathcal{Y}} z_{n,j,y} \overline{R}^d_{y} 
\nonumber \\
& + \sum_{m \in \mathcal{M}} \sum_{t \in \mathcal{T}} \sum_{n \in \mathcal{N}} {\lambda}^\text{ub,m}_{m,t,n,j} \sum_{y \in \mathcal{Y}} z_{n,j,y} \overline{R}^d_{y}
+ \sum_{m \in \mathcal{M}} \sum_{t \in \mathcal{T}} \sum_{n \in \mathcal{N}} T_{t} R^\text{out}_j {\lambda}^\text{out,m}_{m,t,n,j}  \sum_{y \in \mathcal{Y}} z_{n,j,y} \overline{R}^d_{y}
\nonumber \\
&
+ \sum_{m \in \mathcal{M}} \sum_{t \in \mathcal{T}} \sum_{n \in \mathcal{N}}
T_{t} R^\text{in}_j {\lambda}^\text{in,m}_{m,t,n,j} \sum_{y \in \mathcal{Y}} z_{n,j,y} \overline{R}^d_{y}- \sum_{m \in \mathcal{M}} \sum_{t \in \mathcal{T}} \sum_{n \in \mathcal{N}} \sum_{i'\in \mathcal{I'}} \underline{R}_{n,i'} \overline{R}_{n,i'} {\lambda}^\text{lb,p}_{m,t,n,i'}
\nonumber \\
&
+ \sum_{m \in \mathcal{M}} \sum_{t \in \mathcal{T}} \sum_{n \in \mathcal{N}}
\sum_{i'\in \mathcal{I'}} \overline{R}_{n,i'}{\lambda}^\text{ub,p}_{m,t,n,i'}
+ \sum_{m \in \mathcal{M}} \sum_{t \in \mathcal{T}} \sum_{n \in \mathcal{N}} \sum_{i'\in \mathcal{I'}} T_{t} \,  R^\text{out}_{i'} \,  \overline{R}_{n,i'} {\lambda}^\text{out,p}_{m,t,n,i'}
\nonumber \\
&
+ \sum_{m \in \mathcal{M}} \sum_{t \in \mathcal{T}} \sum_{n \in \mathcal{N}}
\sum_{i'\in \mathcal{I'}}  T_{t} \,  R^\text{in}_{i'} \,  \overline{R}_{n,i'}
{\lambda}^\text{in,p}_{m,t,n,i'}+ \sum_{m \in \mathcal{M}} \sum_{t \in \mathcal{T}} \sum_{n \in \mathcal{N}} \sum_{i'\in \mathcal{I'}} \sum_{\substack{e \in \mathcal{E}}} T_{t} \,  A^e_{m,t,n} \, \overline{G}{}^e_{n,i'} {\beta}^{e}_{m,t,n,i'}
\nonumber \\
&
+ \sum_{m \in \mathcal{M}} \sum_{t \in \mathcal{T}} \sum_{n \in \mathcal{N}} \sum_{i'\in \mathcal{I'}} \sum_{\substack{u \in \mathcal{U}_{n,i'}}}
T_{t} \, R^{\text{down}}_u \, \overline{G}^\text{conv}_{n,i',u}{\beta}^{\text{down}}_{m,t,n,i',u} 
\nonumber \\
& + \sum_{m \in \mathcal{M}} \sum_{t \in \mathcal{T}} \sum_{n \in \mathcal{N}}
\sum_{i'\in \mathcal{I'}} \sum_{\substack{u \in \mathcal{U}_{n,i'}}} T_{t} \, R^{\text{up}}_u \, \overline{G}^\text{conv}_{n,i',u} {\beta}^{\text{up}}_{m,t,n,i',u}
\nonumber \\
&
+ \sum_{m \in \mathcal{M}} \sum_{t \in \mathcal{T}} \sum_{n \in \mathcal{N}} \sum_{i'\in \mathcal{I'}} \sum_{\substack{u \in \mathcal{U}_{n,i'}}} T_{t}  \, \overline{G}^\text{conv}_{n,i',u} \beta^\text{conv}_{m,t,n,i',u}.
\end{align}

Now, by isolating the  terms corresponding to the right-hand side of Eq. \eqref{lpduality}, we can rewrite the nonlinear objective function of a merchant operating under Cournot oligopoly \eqref{merchant_upper} as the following quadratic expression\footnote{Under perfect competition, the quadratic objective function of the merchant \eqref{merchant_upper_quadr} lacks the term with the bold font in the second row.}
\begin{align}
\max_{\substack{}} & \Bigg\{  \sum_{m \in \mathcal{M}} \sum_{t \in \mathcal{T}} \sum_{n \in \mathcal{N}} W_{m}   \, \Bigg[  
 D^{\text{int}}_{m,t,n} q_{m,t,n} - 
 D_{m,t,n}^{\text{slp}} q_{m,t,n}^2
\nonumber \\
& \mathbf{ - D_{m,t,n}^{\textbf{slp}} \sum_{i'\in \mathcal{I'}} \bigg( \sum_{u \in \mathcal{U}_{n,i'}} g^\textbf{conv}_{m,t,n,i',u}  + \sum_{\substack{e \in \mathcal{E}}}  g^e_{m,t,n,i'} + r^\textbf{out}_{m,t,n,i'}  - r^\textbf{in}_{m,t,n,i'} \bigg)^2 }
\nonumber \\ 
&  - \sum_{i'\in \mathcal{I'}} \sum_{u \in \mathcal{U}_{n,i'}}  C_{u}^\text{conv}  g^\text{conv}_{m,t,n,i',u}   -  \sum_{i\in \mathcal{I}} C^{\text{sto}} r^{\text{out}}_{m,t,n,i} \Bigg]
\nonumber \\
& - \sum_{m \in \mathcal{M}} \sum_{t \in \mathcal{T}} \sum_{\ell \in \mathcal{L}} T_{t} K_{\ell}  \big( \underline{\mu}_{m,t,\ell} + \overline{\mu}_{m,t,\ell} \big) + \sum_{m \in \mathcal{M}} \sum_{t \in \mathcal{T}} \sum_{n \in \mathcal{N}} \sum_{i'\in \mathcal{I'}} \underline{R}_{n,i'} \overline{R}_{n,i'} {\lambda}^\text{lb,p}_{m,t,n,i'}
\nonumber \\
&
- \sum_{m \in \mathcal{M}} \sum_{t \in \mathcal{T}} \sum_{n \in \mathcal{N}}
\sum_{i'\in \mathcal{I'}} \overline{R}_{n,i'}{\lambda}^\text{ub,p}_{m,t,n,i'}
- \sum_{m \in \mathcal{M}} \sum_{t \in \mathcal{T}} \sum_{n \in \mathcal{N}} \sum_{i'\in \mathcal{I'}} T_{t} \,  R^\text{out}_{i'} \,  \overline{R}_{n,i'} {\lambda}^\text{out,p}_{m,t,n,i'}
\nonumber \\
&
- \sum_{m \in \mathcal{M}} \sum_{t \in \mathcal{T}} \sum_{n \in \mathcal{N}}
\sum_{i'\in \mathcal{I'}}  T_{t} \,  R^\text{in}_{i'} \,  \overline{R}_{n,i'}
{\lambda}^\text{in,p}_{m,t,n,i'}- \sum_{m \in \mathcal{M}} \sum_{t \in \mathcal{T}} \sum_{n \in \mathcal{N}} \sum_{i'\in \mathcal{I'}} \sum_{\substack{e \in \mathcal{E}}} T_{t} \,  A^e_{m,t,n} \, \overline{G}{}^e_{n,i'} {\beta}^{e}_{m,t,n,i'}
\nonumber \\
&
- \sum_{m \in \mathcal{M}} \sum_{t \in \mathcal{T}} \sum_{n \in \mathcal{N}} \sum_{i'\in \mathcal{I'}} \sum_{\substack{u \in \mathcal{U}_{n,i'}}}
T_{t} \, R^{\text{down}}_u \, \overline{G}^\text{conv}_{n,i',u}{\beta}^{\text{down}}_{m,t,n,i',u} \nonumber \\
&- \sum_{m \in \mathcal{M}} \sum_{t \in \mathcal{T}} \sum_{n \in \mathcal{N}}
\sum_{i'\in \mathcal{I'}} \sum_{\substack{u \in \mathcal{U}_{n,i'}}} T_{t} \, R^{\text{up}}_u \, \overline{G}^\text{conv}_{n,i',u} {\beta}^{\text{up}}_{m,t,n,i',u}
\nonumber \\
&
- \sum_{m \in \mathcal{M}} \sum_{t \in \mathcal{T}} \sum_{n \in \mathcal{N}} \sum_{i'\in \mathcal{I'}} \sum_{\substack{u \in \mathcal{U}_{n,i'}}} T_{t}  \, \overline{G}^\text{conv}_{n,i',u} \beta^\text{conv}_{m,t,n,i',u} \Bigg\} - \sum_{n \in \mathcal{N}} \sum_{y \in \mathcal{Y}} z_{n,j,y} I \overline{R}^d_{y}. \label{merchant_upper_quadr}
\end{align}

\subsubsection{Mathematical Program with Primal and Dual Constraints (MPPDC)}

The bi-level problem comprising the upper-level storage investor (either welfare-maximizer or the profit-maximizing merchant) and the lower-level market cannot be solved directly using off-the-shelf mathematical programming tools. In order to compute the market equilibrium, we reformulate the problem into an equivalent single-level problem, relying on the primal-dual approach, MPPDC. Specifically, we take the upper-level objective and respective constraints, and combine those with the primal constraints, dual constraints, and the strong-duality equality of the lower-level ISO problem (i.e., Eq. \eqref{iso_co} s.t. \eqref{prod2}--\eqref{primalvariables4}). As a result, we obtain a mixed integer quadratically constrained quadratic problem (MIQCQP).

The primal constraints are Equations \eqref{prod2}--\eqref{primalvariables4}. The dual constraints of the ISO problem under Cournot oligopoly\footnote{Under perfect competition, dual constraints \eqref{dual_co1}, \eqref{dual_co2}, \eqref{dual_co3}, and \eqref{dual_co4} lack the terms with bold font.} can be written as
\begin{align}
& - W_{m}  \bigg( D^{\text{int}}_{m,t,n} - D_{m,t,n}^{\text{slp}} q_{m,t,n} \bigg) + \theta_{m,t,n} \geq 0 \quad (q_{m,t,n}), \; \forall m,t,n \label{dualfirst}
\\
& - \sum_{\substack{n' \in \mathcal{N}}} T_{t} B_{n,n'} \theta_{m,t,n'} + \sum_{\substack{\ell \in \mathcal{L}}} T_{t}H_{\ell,n} \overline{\mu}_{m,t,\ell}
- \sum_{\substack{\ell \in \mathcal{L}}} T_{t}H_{\ell,n} \underline{\mu}_{m,t,\ell} = 0 \quad (v_{m,t,n}),  \; \forall m,t,n  
\\
&  W_{m}   \bigg( \mathbf{ D_{m,t,n}^{\textbf{slp}} \bigg(  \sum_{u' \in \mathcal{U}_{n,i'}}g^\textbf{conv}_{m,t,n,i',u'} +\sum_{\substack{e \in \mathcal{E}}}g^e_{m,t,n,i'} + r^\textbf{out}_{m,t,n,i'}  - r^\textbf{in}_{m,t,n,i'} \bigg) } + C^{\text{conv}}_u \bigg) 
\nonumber \\
& - \theta_{m,t,n} + \beta^{\text{conv}}_{m,t,n,i',u} + \beta^{\text{up}}_{m,t,n,i',u}  - \beta^{\text{up}}_{m,t+1,n,i',u} + \beta^{\text{down}}_{m,t+1,n,i',u} - \beta^{\text{down}}_{m,t,n,i',u}  \geq 0
\nonumber \\ 
& (g^\text{conv}_{m,t,n,i',u}), \; \forall m,t,n,i',u \in \mathcal{U}_{n,i'}
\label{dual_co1}  \\
& \mathbf{ W_{m} D_{m,t,n}^{\textbf{slp}} \bigg(  \sum_{u' \in \mathcal{U}_{n,i'}}g^\textbf{conv}_{m,t,n,i',u'} +\sum_{\substack{e \in \mathcal{E}}}g^e_{m,t,n,i'} + r^\textbf{out}_{m,t,n,i'}  - r^\textbf{in}_{m,t,n,i'} \bigg) }  - \theta_{m,t,n} + \beta^{e}_{m,t,n,i'} \geq 0 
\nonumber \\ 
& (g^e_{m,t,n,i'}), \; \forall m,t,n,i' \label{dual_co2}
\\
& \lambda^{\text{bal}}_{m,t,n,i'} - (1-E^{\text{sto}}_{i'})^{T_{t}}  \lambda^{\text{bal}}_{m,t+1,n,i'} + \lambda^{\text{ub,p}}_{m,t,n,i'} - \lambda^{\text{lb,p}}_{m,t,n,i'} \geq 0 \quad (r^{\text{sto}}_{m,t,n,i'}), \; \forall m,t,n,i'
\\
 & \lambda^{\text{bal}}_{m,t,n,j} - (1-E^{\text{sto}}_j)^{T_{t}} \lambda^{\text{bal}}_{m,t+1,n,j} + \lambda^{\text{ub,m}}_{m,t,n,j} - \lambda^{\text{lb,m}}_{m,t,n,j} \geq 0 \quad (r^{\text{sto}}_{m,t,n,j}), \; \forall m,t,n,j
\\
& \mathbf{ -  W_{m}   D_{m,t,n}^{\textbf{slp}} \bigg(  \sum_{u' \in \mathcal{U}_{n,i'}}g^\textbf{conv}_{m,t,n,i',u'} +\sum_{\substack{e \in \mathcal{E}}}g^e_{m,t,n,i'} + r^\textbf{out}_{m,t,n,i'}  - r^\textbf{in}_{m,t,n,i'} \bigg) } + \theta_{m,t,n} \nonumber \\
& - E^{\text{in}}_{i'} \lambda^{\text{bal}}_{m,t,n,i'} + \lambda^{\text{in,p}}_{m,t,n,i'}  \geq 0 \quad (r^{\text{in}}_{m,t,n,i'}), \; \forall m,t,n,i' \label{dual_co3}
\\
 & \theta_{m,s,n} - E^{\text{in}}_j \lambda^{\text{bal}}_{m,s,n,j} + \lambda^{\text{in,m}}_{m,s,n,j} \geq 0 \quad (r^{\text{in}}_{m,s,n,j} ), \; \forall m,s,n,j
\\
 &  W_{m}   \bigg( \mathbf{ D_{m,t,n}^{\textbf{slp}} \bigg(  \sum_{u' \in \mathcal{U}_{n,i'}}g^\textbf{conv}_{m,t,n,i',u'} +\sum_{\substack{e \in \mathcal{E}}}g^e_{m,t,n,i'} + r^\textbf{out}_{m,t,n,i'}  - r^\textbf{in}_{m,t,n,i'} \bigg) } + C^{\text{sto}} \bigg) \nonumber \\ 
& - \theta_{m,t,n}  + \lambda^{\text{bal}}_{m,t,n,i'} + \lambda^{\text{out,p}}_{m,t,n,i'}  \geq 0 \quad (r^{\text{out}}_{m,t,n,i'}), \; \forall m,t,n,i'
\label{dual_co4}
\\
 &  W_{m} C^{\text{sto}} - \theta_{m,s,n} + \lambda^{\text{bal}}_{m,s,n,j} + \lambda^{\text{out,m}}_{m,s,n,j}  \geq 0 \quad (r^{\text{out}}_{m,s,n,j}), \; \forall m,s,n,j \label{duallast} .
\end{align}

Instead of writing the strong duality requirements as an equality constraint, we follow the procedure of \cite{Huppmann2015} and write this condition as an inequality to avoid turning the feasible region non-convex due to its quadratic terms. By doing so, the constraint remains convex, but due to strong duality, it will be active as an equality, i.e., when the solution satisfies both primal and dual feasibility conditions. Thus, we obtain:
\begin{align}
& \sum_{m \in \mathcal{M}} \sum_{t \in \mathcal{T}} \sum_{n \in \mathcal{N}} W_{m}   \, \Bigg[  
\bigg( D^{\text{int}}_{m,t,n} q_{m,t,n} - \frac{1}{2} D_{m,t,n}^{\text{slp}} q_{m,t,n}^2 \bigg) 
\nonumber \\
& \mathbf{ - \frac{1}{2} D_{m,t,n}^{\textbf{slp}} \sum_{i'\in \mathcal{I'}} \bigg( \sum_{u \in \mathcal{U}_{n,i'}} g^\textbf{conv}_{m,t,n,i',u}  + \sum_{\substack{e \in \mathcal{E}}}  g^e_{m,t,n,i'} + r^\textbf{out}_{m,t,n,i'}  - r^\textbf{in}_{m,t,n,i'} \bigg)^2 }
\nonumber \\ 
&  - \sum_{i'\in \mathcal{I'}} \sum_{u \in \mathcal{U}_{n,i'}}  C_{u}^\text{conv}  g^\text{conv}_{m,t,n,i',u}   -  \sum_{i\in \mathcal{I}} C^{\text{sto}} r^{\text{out}}_{m,t,n,i} \Bigg]
\nonumber \\
\ge & \sum_{m \in \mathcal{M}} \sum_{t \in \mathcal{T}} \sum_{n \in \mathcal{N}} W_{m}  \, \Bigg[  
 \frac{1}{2} D_{m,t,n}^{\text{slp}} \bigg(  q_{m,t,n}^2 + \mathbf{ \sum_{i'\in \mathcal{I'}} \bigg( \sum_{u \in \mathcal{U}_{n,i'}} g^\textbf{conv}_{m,t,n,i',u}  + \sum_{\substack{e \in \mathcal{E}}}  g^e_{m,t,n,i'} }
 \nonumber \\ 
& \mathbf{ + r^\textbf{out}_{m,t,n,i'}  - r^\textbf{in}_{m,t,n,i'} \bigg)^2  \bigg) }
 \Bigg] 
\nonumber \\
& + \sum_{m \in \mathcal{M}} \sum_{t \in \mathcal{T}} \sum_{\ell \in \mathcal{L}} T_{t} K_{\ell}  \big( \underline{\mu}_{m,t,\ell} + \overline{\mu}_{m,t,\ell} \big) - \sum_{m \in \mathcal{M}} \sum_{t \in \mathcal{T}} \sum_{n \in \mathcal{N}} \underline{R}_{n,j} {\lambda}^\text{lb,m}_{m,t,n,j} \sum_{y \in \mathcal{Y}} z_{n,j,y} \overline{R}^d_{y} 
\nonumber \\
& + \sum_{m \in \mathcal{M}} \sum_{t \in \mathcal{T}} \sum_{n \in \mathcal{N}} {\lambda}^\text{ub,m}_{m,t,n,j} \sum_{y \in \mathcal{Y}} z_{n,j,y} \overline{R}^d_{y}
+ \sum_{m \in \mathcal{M}} \sum_{t \in \mathcal{T}} \sum_{n \in \mathcal{N}} T_{t} R^\text{out}_j {\lambda}^\text{out,m}_{m,t,n,j}  \sum_{y \in \mathcal{Y}} z_{n,j,y} \overline{R}^d_{y}
\nonumber \\
&
+ \sum_{m \in \mathcal{M}} \sum_{t \in \mathcal{T}} \sum_{n \in \mathcal{N}}
T_{t} R^\text{in}_j {\lambda}^\text{in,m}_{m,t,n,j} \sum_{y \in \mathcal{Y}} z_{n,j,y} \overline{R}^d_{y}
- \sum_{m \in \mathcal{M}} \sum_{t \in \mathcal{T}} \sum_{n \in \mathcal{N}} \sum_{i'\in \mathcal{I'}} \underline{R}_{n,i'} \overline{R}_{n,i'} {\lambda}^\text{lb,p}_{m,t,n,i'}
\nonumber \\
&
+ \sum_{m \in \mathcal{M}} \sum_{t \in \mathcal{T}} \sum_{n \in \mathcal{N}}
\sum_{i'\in \mathcal{I'}} \overline{R}_{n,i'}{\lambda}^\text{ub,p}_{m,t,n,i'}
+ \sum_{m \in \mathcal{M}} \sum_{t \in \mathcal{T}} \sum_{n \in \mathcal{N}} \sum_{i'\in \mathcal{I'}} T_{t} \,  R^\text{out}_{i'} \,  \overline{R}_{n,i'} {\lambda}^\text{out,p}_{m,t,n,i'}
\nonumber \\
&
+ \sum_{m \in \mathcal{M}} \sum_{t \in \mathcal{T}} \sum_{n \in \mathcal{N}}
\sum_{i'\in \mathcal{I'}}  T_{t} \,  R^\text{in}_{i'} \,  \overline{R}_{n,i'}
{\lambda}^\text{in,p}_{m,t,n,i'}+ \sum_{m \in \mathcal{M}} \sum_{t \in \mathcal{T}} \sum_{n \in \mathcal{N}} \sum_{i'\in \mathcal{I'}} \sum_{\substack{e \in \mathcal{E}}} T_{t} \,  A^e_{m,t,n} \, \overline{G}{}^e_{n,i'} {\beta}^{e}_{m,t,n,i'}
\nonumber \\
&
+ \sum_{m \in \mathcal{M}} \sum_{t \in \mathcal{T}} \sum_{n \in \mathcal{N}} \sum_{i'\in \mathcal{I'}} \sum_{\substack{u \in \mathcal{U}_{n,i'}}}
T_{t} \, R^{\text{down}}_u \, \overline{G}^\text{conv}_{n,i',u}{\beta}^{\text{down}}_{m,t,n,i',u} \nonumber \\
&+ \sum_{m \in \mathcal{M}} \sum_{t \in \mathcal{T}} \sum_{n \in \mathcal{N}}
\sum_{i'\in \mathcal{I'}} \sum_{\substack{u \in \mathcal{U}_{n,i'}}} T_{t} \, R^{\text{up}}_u \, \overline{G}^\text{conv}_{n,i',u} {\beta}^{\text{up}}_{m,t,n,i',u}
\nonumber \\
&
+ \sum_{m \in \mathcal{M}} \sum_{t \in \mathcal{T}} \sum_{n \in \mathcal{N}} \sum_{i'\in \mathcal{I'}} \sum_{\substack{u \in \mathcal{U}_{n,i'}}} T_{t}  \, \overline{G}^\text{conv}_{n,i',u} \beta^\text{conv}_{m,t,n,i',u}. \label{SDE_CO}
\end{align}

The strong-duality equality \eqref{SDE_CO} includes four nonlinear terms as multiplications of the storage investor's dual variables ($\lambda^\text{lb,m}_{m,t,n,j}$, $\lambda^\text{ub,m}_{m,t,n,j}$, $\lambda^\text{out,m}_{m,t,n,j}$,  $\lambda^\text{in,m}_{m,t,n,j}$) and the discrete storage investment decision ($\sum_{y \in \mathcal{Y}} z_{n,j,y} \overline{R}^d_{y}$). We present the exact linearization applied to them in Appendix \ref{appendix_a}.

The resulting mathematical formulations as MIQCQPs are summarized in Table \ref{table:models}. The central-planning benchmark model is abbreviated as CP. Other abbreviations denote ``upper-level'' -- ``lower-level'' with ``SW'' for a welfare-maximizing storage investor and ``M'' for the merchant storage investor. ``PC'' stands for perfect competition and ``CO'' for Cournot oligopoly at the lower-level market.

\begin{table}[!ht]
\caption{MIQCQP formulations: For 2.-4., the equations are the objective function subject to the upper level constraint, and lower-level constraints as primal constraints, dual constraints, the strong-duality equality, and constraints needed to linearize the strong-duality expression, respectively. * denotes the PC version: please refer to the footnotes of the corresponding constraints.}
\label{table:models}
\centering
\begin{tabular}{l|l}
     Model (upper level - lower level) & Mathematical formulation \\
     \hline
     \hline
     1. CP (benchmark) & \eqref{swmax_cp} s.t. \eqref{binary_constraint}-\eqref{primalvariables4}  \\
     \hline
     \hline
     2. SW-PC & \eqref{swmax_upper} s.t. \eqref{binary_constraint}-\eqref{primalvariables4}, \eqref{dualfirst}*-\eqref{duallast}*, \eqref{SDE_CO_2}*, \eqref{linearization_first}-\eqref{linearization_last}   \\
     \hline
     3. M-PC & \eqref{merchant_upper_quadr}* s.t. \eqref{binary_constraint}-\eqref{primalvariables4}, \eqref{dualfirst}*-\eqref{duallast}*, \eqref{SDE_CO_2}*, \eqref{linearization_first}-\eqref{linearization_last}  \\
     \hline
     4. SW-CO & \eqref{swmax_upper} s.t. \eqref{binary_constraint}-\eqref{primalvariables4}, \eqref{dualfirst}-\eqref{duallast}, \eqref{SDE_CO_2}, \eqref{linearization_first}-\eqref{linearization_last}  \\
     \hline
     5. M-CO & \eqref{merchant_upper_quadr} s.t. \eqref{binary_constraint}-\eqref{primalvariables4}, \eqref{dualfirst}-\eqref{duallast}, \eqref{SDE_CO_2}, \eqref{linearization_first}-\eqref{linearization_last}  \\
\end{tabular}
\end{table}

\section{Numerical Examples} \label{numerical_examples}

\subsection{Illustrative Example: Three-Node Network} \label{3node}

\begin{figure}[h]
  \centering
  \includegraphics[width=8cm]{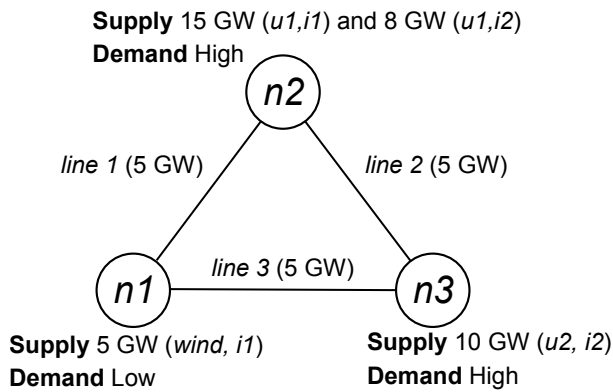}
  \caption{Network and data for the three-node example.}
  \label{fig:3nodes}
\end{figure}

The illustrative example for three nodes is presented in Figure \ref{fig:3nodes}. All three transmission lines are identical. We consider a simple two-period example with an off-peak ($t1$) and a peak ($t2$) hour. We also incorporate two seasons of equal weights. These are characterized as high demand - low wind ($m1$), and low demand  - high wind ($m2$). There is one investor who chooses the investment locations and sizes between 0 and 200 MWh with 50 MWh increments in each node. There is no existing storage capacity, and new capacity is 100\% efficient with a 50\% charge rate. Other parameters are chosen with the aim of producing qualitative insights: two producers $i1$ and $i2$ with VRES (wind) and two available conventional technologies: cheap but relatively inflexible $u1$ and more expensive but flexible $u2$. Installed capacities are presented in Fig. \ref{fig:3nodes}.\footnote{All problem instances implemented as either MIQPs or MIQCQPs solve to optimality with CPLEX 12.8 in a matter of seconds with GAMS 26.1 and a workstation with 2.40 GHz Intel i5-6300 processor and 8 GB of RAM.}

Table \ref{table:3node_results} summarizes illustrative results. First, we note that the central-planning equilibrium (1. CP) equals the one with a welfare-maximizer under perfect competition (2. SW-PC). This is expected because the two models effectively represent the same objectives. The optimal decision is to invest in full capacity in all nodes. This increases social welfare (SW) by 0.03\% and decreases grid revenue (GR), the value of imports, by 4\% compared to PC without any investments. A merchant under PC (3. M-PC) benefits from investing less. It has 200 MWh in $n1$ and $n3$, but only 50 MWh in $n2$, which acts as a net exporter and has a smaller price gap between the off-peak and peak hours in the high demand - low wind season ($m1$). Both investors refrain from making any investment under Cournot oligopoly (4. and 5.) with an uncongested network.

\begin{table}[!h]
\scriptsize
\caption{Results for the illustrative three-node example: social welfare (SW), investor surplus (IS), producer surplus (PS), consumer surplus (CS), grid revenue (GR), equilibrium prices and storage investments. }
\centering
\begin{tabular}{l|l|l|l|l|l|l|l}
\label{table:3node_results}
     Model & SW (k\euro) & IS (k\euro) & PS (k\euro) & CS (k\euro) & GR (k\euro) & Price (\euro/MWh) & Investment (GWh) \\
         \hline
         \hline
     1. CP          & 6 514.15 & 1.30 & 2 178.56 & 4 241.81 & 92.48 & 60.54 & 0.2 ($n1$), 0.2 ($n2$), 0.2 ($n3$) \\
         \hline
         \hline
     2. SW-PC       & 6 514.15 & 1.30 & 2 178.56 & 4 241.81 & 92.48 & 60.54 & 0.2 ($n1$), 0.2 ($n2$), 0.2 ($n3$) \\
         \hline
     3. M-PC & 6 514.07 & 1.38 & 2 168.54 & 4 251.01 & 93.14 & 60.36 & 0.2 ($n1$), 0.05 ($n2$), 0.2 ($n3$) \\
         \hline
     4. SW-CO       & 5 471.15 & -	  & 3 496.42 & 1 974.73 & -	& 103.10 & -  \\
         \hline
     5. M-CO & 5 471.15 & -	  & 3 496.42 & 1 974.73 & -	& 103.10 & -  \\       
\end{tabular}
\end{table}

\subsection{Case Study: Western European Network} 
\label{fullmodel}

\subsubsection{Numerical Issues with Large-Scale MIQCQPs} \label{numerical_issues}

Although we are able to solve the presented MIQCQP formulations with smaller instances, such as the presented three-node example (Fig. \ref{fig:3nodes}), solving them for more realistic large-scale instances is significantly more challenging. In fact, state-of-the-art solvers available in GAMS do not appear to be able to solve the case study setting presented later in this section credibly and consistently even with several simplifications to reduce the instance size. Specifically, we tested the simplest MIQCQP formulation (2. SW-PC) with a decreased time resolution of 4-hour blocks ($T_t$=4 h). CPLEX 12.8, even with extensive solver setting testing, often reported finding a ``proven optimal solution'', although the strong-duality equality of Eq. \eqref{SDE_CO_2} not being satisfied. A solvable size for CPLEX seems to be around 10-20 k variables and constraints; the full model is over 1 million. Out of the tested solvers, KNITRO 11.1 seems to perform the best, by being able to solve a reduced version of SW-PC with ca. 200 k variables and constraints. However, this did not transfer to other types of the models (3.-5.), and seems to be sensitive to the instance size: usually, the solver terminates with infeasibility\footnote{Terminate at infeasible point because the relative change in solution estimate $<$ 1.0e-15 (default tolerance) for three consecutive iterations.} or primal feasibility\footnote{Primal feasible solution; terminate because the relative change in the objective function $<$ 1.0e-15 for five consecutive feasible iterations.}, or does not progress within the time limit. GUROBI 8.1 returns a suboptimal solution even for the reduced model of 200 k variables and constraints. BONMINH 1.8, BARON 18.5, and MOSEK 8.1 did not show promising results even after 8 hours for the full model, so they were not tested further. The issues stem from computational difficulties caused by the quadratic constraint (strong-duality expression, Eq. \eqref{SDE_CO_2}).

Finally, an alternative approach to reformulate the initial bi-level problem would be to build an equivalent single-level mathematical program with equilibrium constraints (MPEC) by using disjunctive constraints from the lower-level's  first order conditions for optimality (Karush-Kuhn-Tucker, KKT, conditions) \citep{GL2010,GCFHR2012}. Nevertheless, this approach proved intractable due to the large number of binary variables that it required for our problem instance. The complexity of the model, size of the instance (i.e., the number of producers, studied hours, nodes in the network, etc.), and the inclusion of storage operations quickly rendered computationally intractable problems.

\subsubsection{Iterative QP-Based Method} \label{problemformulation_alternative}

To overcome the numerical issues with solving large-scale instances of MIQCQPs, we employed a method that exploits the possibility of enumerating meaningful upper-level decisions. This allows us to study their impacts on the lower level \textit{ex post} and work our way backwards by analyzing which decisions yield the optimal outcome for each decision maker. In practice, we iterate the lower-level market model for the ISO (\eqref{iso_co} s.t. \eqref{prod2}--\eqref{primalvariables4}, either PC or CO), a manageable quadratic programming problem (QP), for all possible size-location investment options. Hence, we make an exhaustive search and now that the optimal investments are unique. The problem instances are implemented in GAMS 26.1 and solved with CPLEX 12.8.\footnote{On a workstation with 2.40 GHz Intel i5-6300 processor and 8 GB of RAM the computational time for a single lower-level model with CPLEX is ca. 10-20 seconds with an additional 30-40 seconds for compilation and post-processing. Therefore, an iteration over $2^7=128$ combinations (i.e., two available investment options for seven possible locations, see more details in Section \ref{data}) takes less than 2 hours.}

\subsubsection{Input Data} \label{data}

\paragraph{Nodes and Transmission Lines}

\begin{figure}[ht!]
  \centering
  \includegraphics[width=8cm]{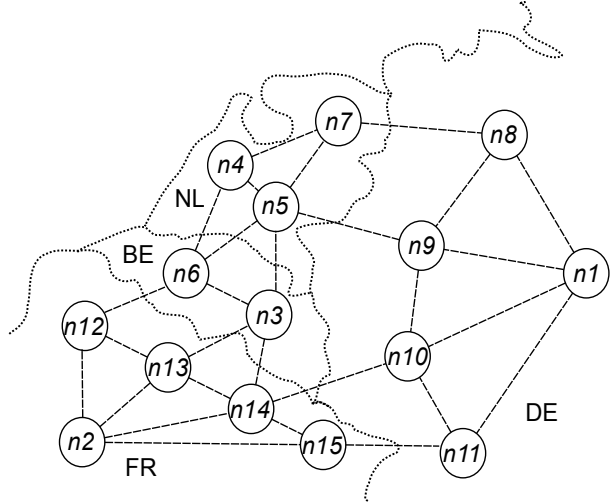}
  \caption{Stylized Western European test network.}
  \label{fig:nodes}
\end{figure}

We use the Western European test network introduced by \cite{Neuhoff2005} and \cite{GL2010} (Fig. \ref{fig:nodes}), where nodes $n1-n7$ represent actual nodes with demand, while $n8-n15$ are auxiliary for modeling transmission. This also extends the case study in \cite{Virasjoki2016}.

\paragraph{Generation and Storage Technologies}

\begin{table}[!h]
\footnotesize
\caption{Marginal costs (with an assumed CO$_2$ price of 20\euro/t), maximum hourly ramping rate, and CO$_2$ emissions for conventional generation.}
\label{table:costs}
\centering
\begin{tabular}{l|c|c|c}
Type & Marginal cost & Max hourly  & CO$_2$ emissions per  \\
& (\euro/MWh) & ramping rate (\%)  & unit of electricity \\
&  &    & output (kg/kWh))\\
\hline
\hline
$u1$ (nuclear) & 9  & 10 & 0    \\
$u2$ (lignite) & 30 & 10 & 0.94 \\
$u3$ (coal)    & 44 & 20 & 0.83 \\
$u4$ (CCGT)    & 39 & 30 & 0.37 \\
$u5$ (gas)     & 53 & 30 & 0.50 \\
$u6$ (oil)     & 91 & 70 & 0.72 \\
$u7$ (hydro)   & 0  & 30 & 0    \\
\end{tabular}
\end{table}

\begin{table}[!h]
\footnotesize
\caption{Estimated installed generation capacity (GW) of the largest producers  in 2017 and used availability percentages per conventional technology $u1-u7$, solar (S), and wind (W) power. }
\label{table:capacity}
\centering
\begin{tabular}{>{\centering\arraybackslash}m{0.8cm} m{2.9cm}|c c c c c c c >{\centering\arraybackslash}m{1.4cm} >{\centering\arraybackslash}m{1.4cm}}
Node & Producer      &$u1$&$u2$&$u3$&$u4$&$u5$&$u6$&$u7$ & S  & W \\
\hline
\hline
$n1$   & Uniper      & -  &0.9 &3.2 &2.7 &0.5 &1.2 & -   & -  & 0.3  \\
       & RWE         &2.6 &9.1 &2.8 &2.5 &1.7 & -  & 0.3 & -  & 0.3  \\
       & EnBW        &2.7 &0.9 &3.0 &0.4 & -  &0.4 & 0.2 & -  & 0.3  \\
       & Vattenfall  & -  & -  &2.9 &0.6 &0.9 &0.1 & -   & -  & 0.6  \\
       & FringeD     & 4.2& 7.4&9.3 &10.9&2.2 &0.4 & 1.3 &40.1& 54.6 \\
\hline
$n2$   & EDF         &63.1& -  &4.0 &1.4 & -  &7.0 &15.0 &0.3 & 1.5  \\
       & FringeF     & -  & -  & -  &3.8 &2.4 & -  &3.6  &6.5 & 12.3  \\
\hline
$n3$,  & Electrabel  &5.9 & -  & -  &1.7 &1.4 & -  & -   & -  & 0.5  \\
$n6$  & EDF (Luminus)& -  & -  & -  &0.4 &0.4 & -  & -   & -  & 0.2  \\
       & FringeB     & -  & -  & -  &1.0 & -  & -  & -   &3.3 & 2.2  \\
\hline
$n4$,  & Electrabel  & -  & -  & -  &2.8 &0.1 & -  & -   & -  & -   \\
$n5$,  & Essent/RWE  & -  & -  &1.3 &1.9 &0.6 & -  & -   & -  & -    \\
$n7$&Nuon/Vattenfall & -  & -  &0.9 &3.2 &1.1 & -  & -   & -  & -   \\
       & FringeN     &0.5 & -  &2.9 &3.2 &0.7 & -  & -   &2.0 &4.3  \\
\hline
\hline
$\%$   & Available   & 80 & 85 & 84 & 89 & 86 & 86 & 30 & Fig. \ref{fig:demand_solar} & Fig. \ref{fig:demand_wind} \\
\end{tabular}
\end{table}

The generation and cost data are updated to correspond to 2017 (Tables \ref{table:costs} and \ref{table:capacity}), although the network is assumed to remain unchanged. Price elasticity is assumed to be -0.25. Operational storage capacity estimates for 2017 in GWh are based on installed power capacities (GW) as reported in the database by \cite{SANDIA}.\footnote{$n1$: 7 GWh (Uniper), 11 GWh (RWE), 1 GWh (EnBW), 17 GWh (Vattenfall), 10 (Fringe); $n2$: 34 GWh (EDF); $n3$ and $n6$: 1 and 5 GWh (Electrabel.) \label{storagefootnote}} Because most of the existing storage capacity is pumped hydro storage, we assume a round-trip efficiency of $E^{\text{in}}_{i'}$ = 0.75, and charge and discharge efficiencies of $R^{\text{in}}_{i'}$ = $R^{\text{out}}_{i'}$ = 0.16. Decay is assumed to be 0\%, in addition to zero marginal costs of using storage ($C^{\text{sto}}$). 

The technology for storage investment is assumed to be utility-scale battery storage, such as lithium-ion storage. We, therefore, use  $E^{\text{in}}_{j}$ = 0.95 and  $R^{\text{in}}_{j}$ = $R^{\text{out}}_{j}$ = 0.50, while decay and operating costs remain the same as existing storage. For the amortized investment cost for one week, we use low-range estimates from \euro80/MWh downward and choose \euro50/MWh as a base case.\footnote{This is based on a total lifetime of 20 years and the operational time frame of one week for the purpose of this case study. Low-range estimates are justifiable since lithium-ion technology costs are expected to have a decreasing trend along with increasing installment rates \citep{IRENA2017}. 
Similar to us, \cite{Dvorkin2018} consider only day-ahead markets: even in the low-cost scenario \$120/kWh only 7/240 nodes are profitable for storage. Therefore, we use similar low-range cost for the economic interest of our case study.} Total storage investment per each investor-node combination can be selected from a set of two discrete options: 0 MWh or 100 MWh. 

\paragraph{Clustering: Representative Weeks}

VRES production and power demand are typically seasonal and include uncertainties, which affect storage installment decisions. However, modeling a long time frame, such as an entire year, is not computationally tractable. One approach to tackle this in a reduced time frame is to use representative days (see, e.g., \cite{Nahmmacher2016,Reichenberg2018}). We use the hierarchical clustering method and model developed by \cite{Reichenberg2019} to create representative weeks based on the 2017 data. The data are hourly (8760 h), and we use time series for demand, wind power production, and solar PV power production for each of the four countries. The data are normalized so that demand in each region obtains values between 0 and 1, where 1 corresponds to the maximum regional demand. For VRES production, the maximum values for normalization are based on the installed capacities, meaning that the normalized data never reach 1. There is no weighting between the time series, e.g., based on their relative importance or the market sizes. Finally, the obtained representative weeks are real weeks that best describe a cluster of similar ones, i.e., are closest to the centroid of the cluster. For more details on the definition of representative weeks, please refer to \cite{Reichenberg2019}. 



\begin{figure}
\centering
\subfigure[Demand - Wind Production]{\label{fig:demand_wind}\includegraphics[width=85mm]{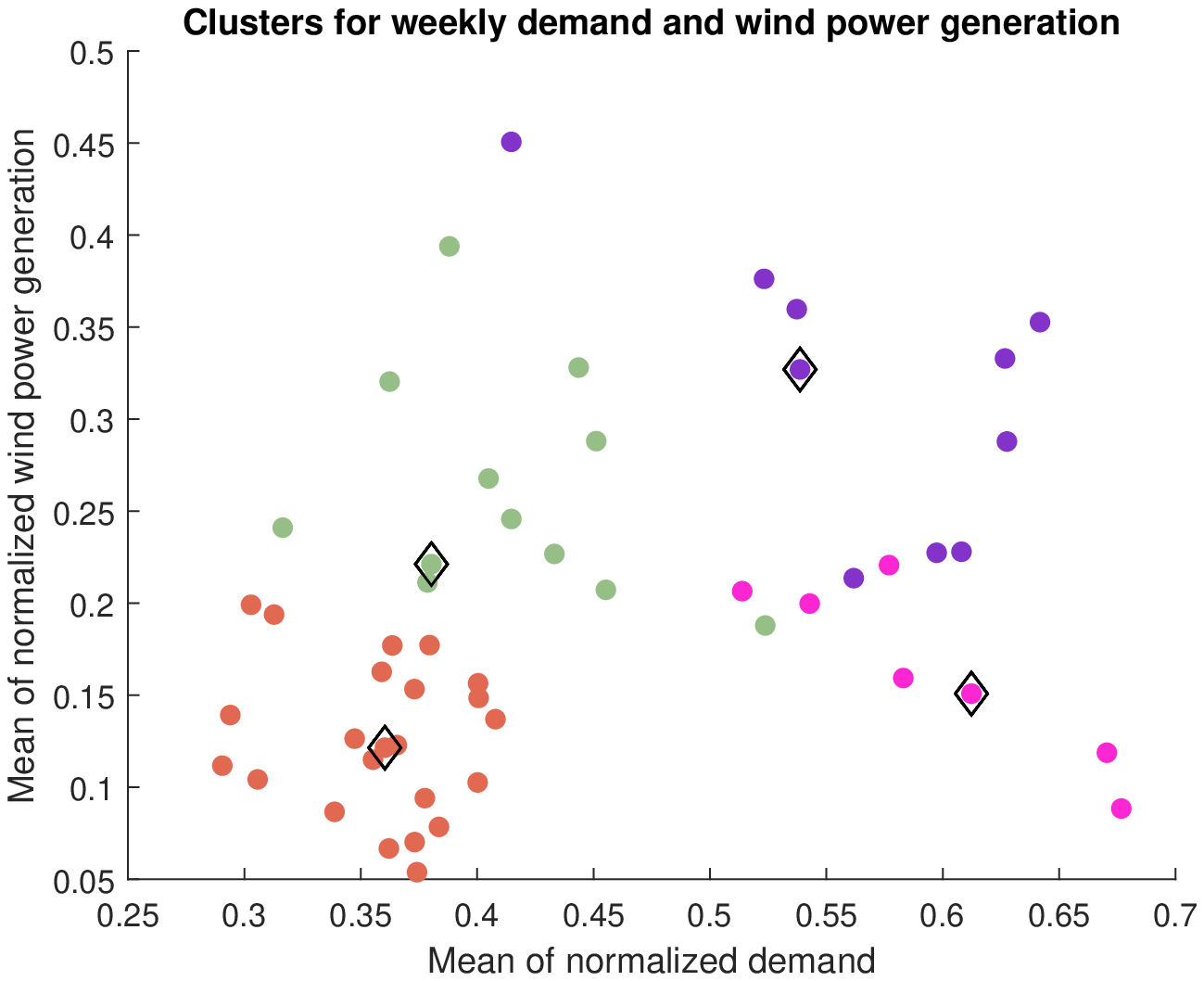}}
\subfigure[Demand - Solar PV Production]{\label{fig:demand_solar}\includegraphics[width=85mm]{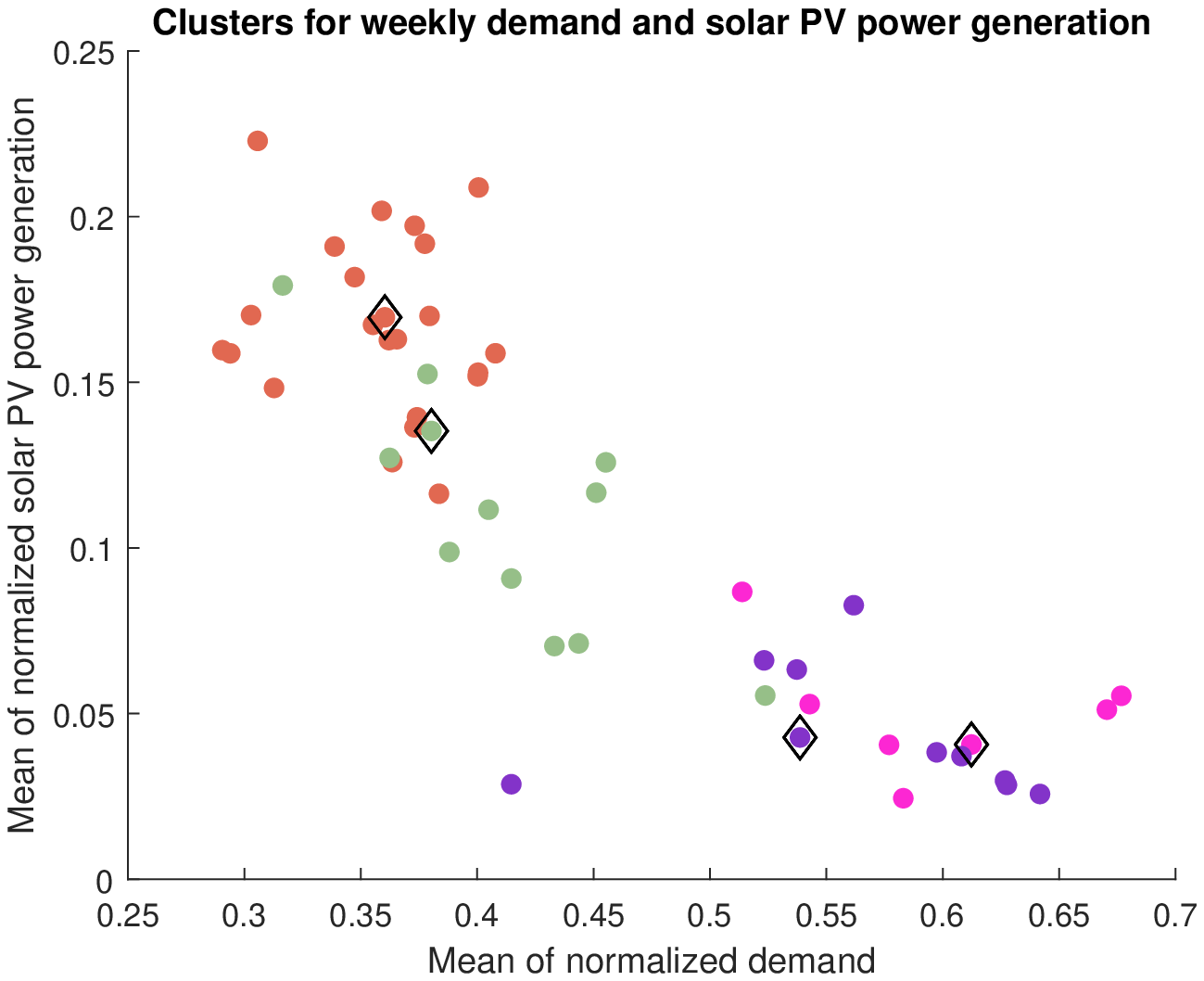}}
\caption{Weeks within the clusters based on the means of normalized demand and either \ref{fig:demand_wind} wind or \ref{fig:demand_solar} solar PV power generation. Representative weeks are denoted by diamonds: week 6 (pink), week 18 (green), week 20 (orange), week 47 (purple).}
\end{figure}

For computational reasons, we choose to use four representative weeks with hourly data ($T_t$ = 1 h) hereinafter.\footnote{Under central-planning, both social welfare and storage investment size remain similar with 4, 6, 8, 15, 20, and 25 weeks, where the lowest social welfare (4 weeks) is 7\% smaller than the highest (25 weeks).} These four weeks $m$ selected by the clustering method are detailed in Table \ref{table:repr_weeks} and Figures \ref{fig:demand_wind} and \ref{fig:demand_solar}. The weights of these weeks ($W_m$) are based on the number of weeks belonging to the cluster. Furthermore, Table \ref{table:repr_weeks} describes the weeks based on their demand, wind, and solar PV power production profiles. In particular, high demand occurs when solar PV production is low during the colder season (weeks 6 and 47) and vice versa for the warmer season (weeks 18 and 20). However, based on the data being used, both seasons can have either high (weeks 18 and 47) or low (weeks 6 and 20) wind production profile.

\begin{table}[!h]
\caption{Representative weeks selected with the clustering method.}
\centering
\small
\begin{tabular}{c|c|c|c|c}
\label{table:repr_weeks}
     Week, $m$ (1-52) & Weight, $W_m$ & Demand profile & Wind profile & Solar profile  \\
     \hline
     \hline
     6  & 7/52  & high & low  & low \\
     18 & 12/52 & low  & high & high \\
     20 & 23/52 & low  & low  & high \\
     47 & 10/52 & high & high & low \\
\end{tabular}
\end{table}

\subsection{Results} \label{results}

\subsubsection{Storage Investment Sizes and Locations}

A welfare-maximizing investor operating in a perfectly competitive market (2. SW-PC) yields effectively the same market equilibrium as under a central-planning perspective, as previously observed in the three-node example in Section \ref{3node}. For the Western European case study, the investor's optimal strategy is to invest 100 MWh in nodes $n2$ (France), together with $n3$ and $n6$ (Belgium), while it chooses not to invest in the remainder of the system. This totals 300 MWh of new investments in countries with a generation mix dominated by nuclear power. In fact, both France and Belgium already have storage, but the Netherlands ($n4$, $n5$, $n7$) has no installed storage capacity. However, the Netherlands enjoys enough flexibility with coal and gas plants along with sufficient interconnections to neighboring countries. Additionally, France and Belgium are more profitable for storage investments due to higher prices and higher price volatility than the rest of the system. This generates more temporal arbitrage opportunities in $n2$, $n3$, and $n6$. 

The merchant's optimal investment strategy (3. M-PC) does not differ from that of a welfare-maximizer under perfect competition. Its profit is maximized with 100 MWh in $n2$, $n3$, and $n6$ each. When the competition in the lower-level market is imperfect, total investments in the studied network are significantly lower. A welfare-maximizer under Cournot oligopoly (4. SW-CO) invests 100 MWh only in $n1$ (Germany). Again, the merchant's profit-maximizing decision (5. M-CO) is identical. Clearly, the underlying market dynamics, specifically the equilibrium prices and their temporal and spatial variations, affect optimal investment decisions more than the different objectives of the investor.

\subsubsection{Welfare Effects}

\begin{table}[!h]
\scriptsize
\caption{Market equilibrium under perfect competition (PC) or Cournot oligopoly (CO) indicated by social welfare (SW), investor surplus (IS), producer surplus (PS), consumer surplus (CS), and grid revenue (GR) (value of imports). Results denote absolute changes to a market without any investments (+/-). }
\centering
\begin{tabular}{c|c|l|l|l|l|l|l}
\label{table:welfare_effects_basecase}
     Amortized investment & Model & SW (k\euro) & IS (k\euro) & PS (k\euro) & CS (k\euro) & GR (k\euro) & Investment, \\
     cost (\euro/MWh) &&&&&&& (GWh) \\
         \hline
         \hline
     \textit{no investment} &  PC & 2 201 628.15 & - &  438 683.35
 & 1 749 290.80 & 13 654.01  & -   \\
         \hline
     50 &  1. CP / 2. SW-PC & +7.04 & +6.45 & -86.53 & +88.02 & -0.90  & +0.3 \\
        &  3. M-PC   & +7.04 & +6.45 & -86.53 & +88.02 & -0.90  & +0.3  \\ 
       \hline
       \hline
     \textit{no investment} &  CO & 1 923 769.99 & - &  989 457.19   
 & 917 432.27 &  16 880.53  & -   \\
       \hline
     50 &  4. SW-CO         & +0.66 & +0.10 & +0.56 & +0.60 & -0.60  & +0.1 \\
        &  5. M-CO   & +0.66 & +0.10 & +0.56 & +0.60 & -0.60  & +0.1  \\          
\end{tabular}
\end{table}

Table \ref{table:welfare_effects_basecase} shows detailed impacts on the market equilibrium with each investor's optimal investments. Most prominently, the investments increase social welfare (SW). Under perfect competition, this actually hampers producer surplus (PS) more than the investor surplus (IS) increases, making consumers (CS) the ultimate beneficiaries, even if they play no direct part in the investment decision. New storage installments decrease the grid revenue of the ISO, which is calculated as the value of imports. Welfare effects under Cournot oligopoly are similar, except for producers, who can also benefit from new installments. However, the effects are significantly smaller, presumably because the investment sizes are smaller and prices are flatter.

\subsubsection{Sensitivity Analysis}

\paragraph{Amortized Investment Cost}

\begin{figure}[!h]
\centering
\subfigure[Welfare-maximizing investor]{\label{fig:sw_pc_inv}\includegraphics[width=80mm]{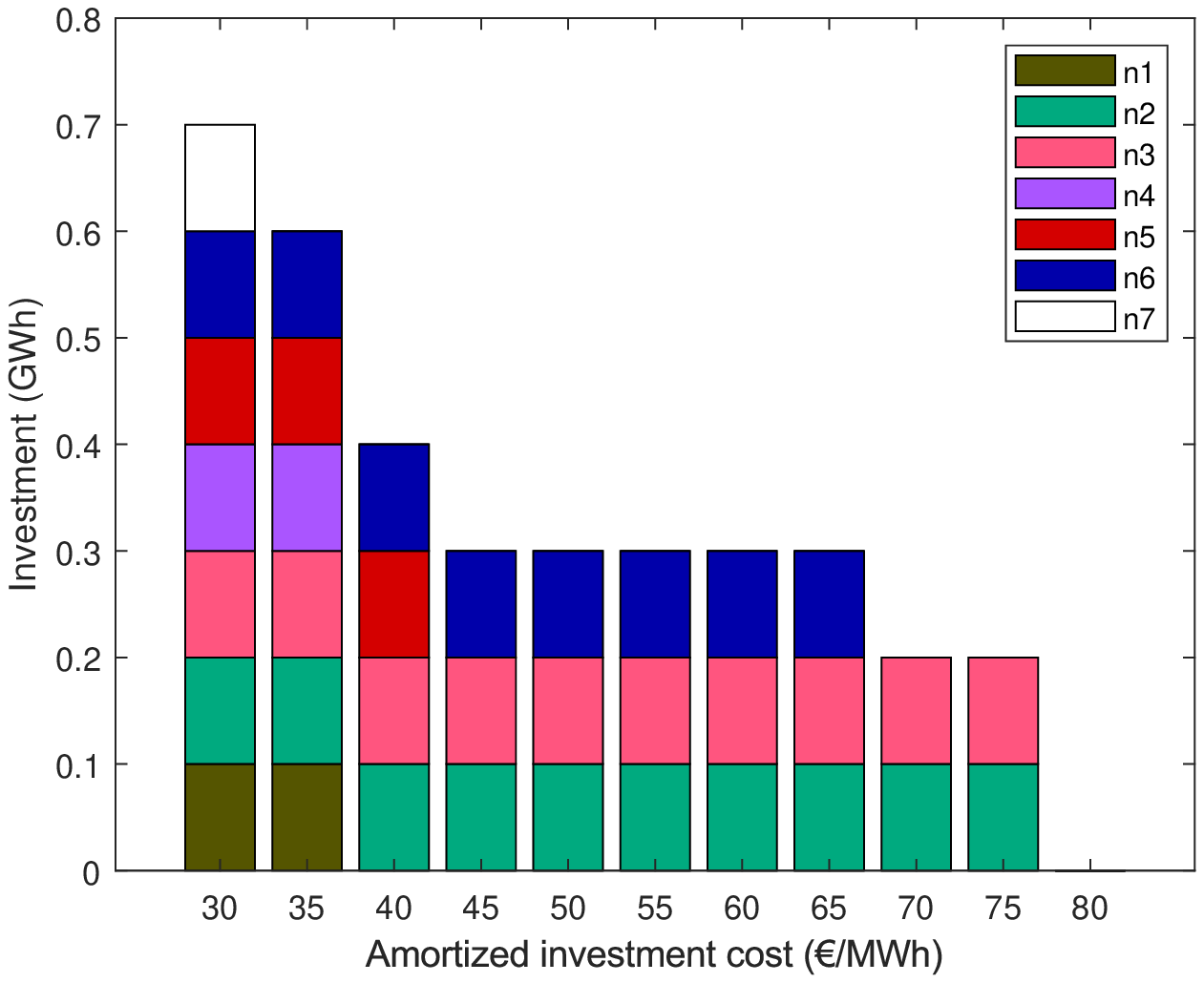}}
\subfigure[Profit-maximizing merchant]{\label{fig:me_pc_inv}\includegraphics[width=80mm]{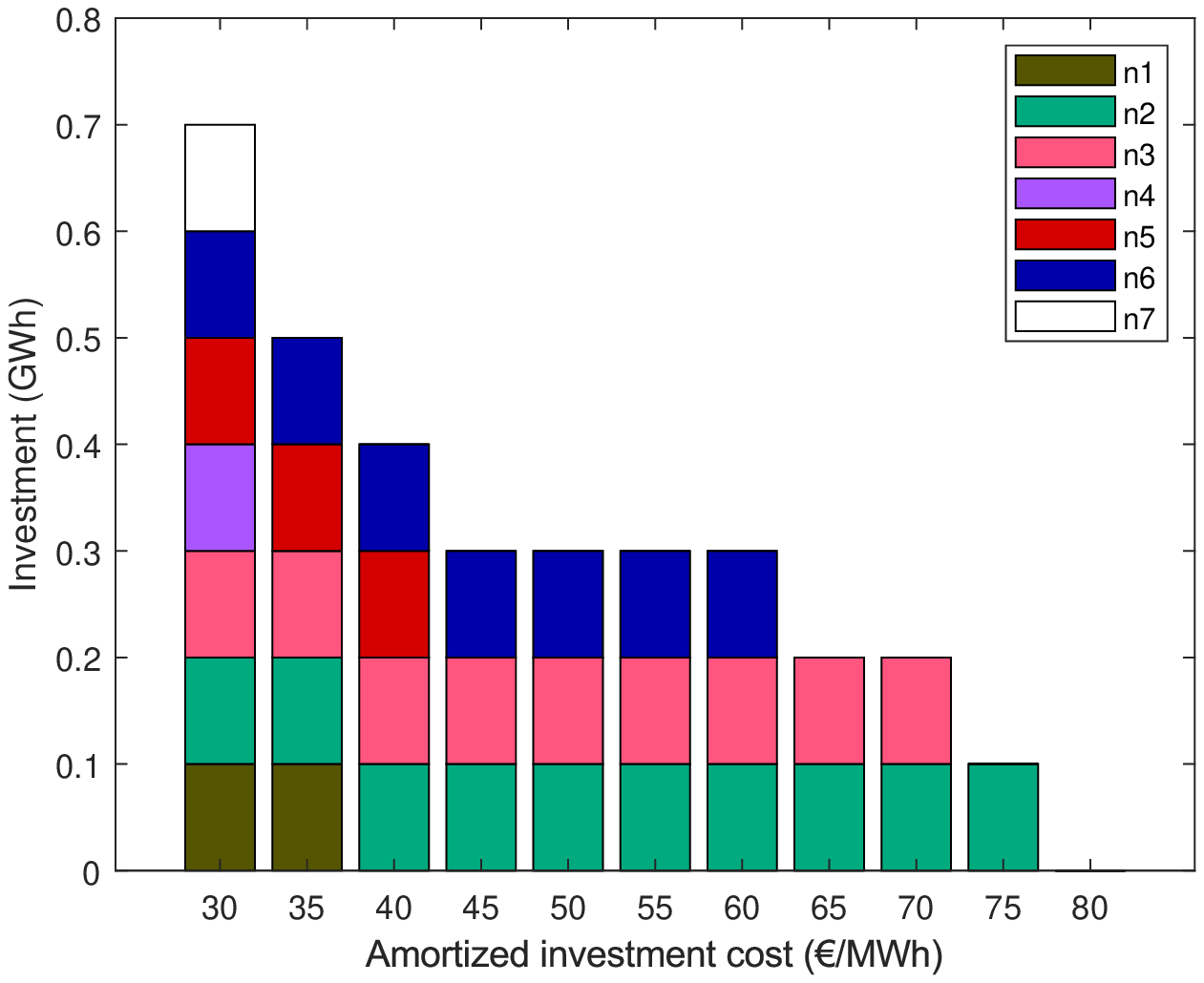}}
\caption{Optimal storage investment sizes and locations under perfect competition}
\label{fig:pc_investments}
\end{figure}

Under perfect competition, new storage investments become profitable at amortized investment cost of \euro80/MWh and lower (Fig. \ref{fig:pc_investments}). The welfare-maximizer (Fig. \ref{fig:sw_pc_inv}) and merchant (Fig. \ref{fig:me_pc_inv}) invest at similar nodes but with a slightly different cost threshold. The investments of the welfare-maximizer are always greater than or equal to the total investments of a merchant. When the investment decisions differ, there is less power plant ramping in the system with the higher investment of the welfare-maximizer. Therefore, the increase in technical flexibility results in a higher economic efficiency.

\begin{figure}[!h]
\centering
\subfigure[Welfare-maximizing investor]{\label{fig:sw_co_inv}\includegraphics[width=80mm]{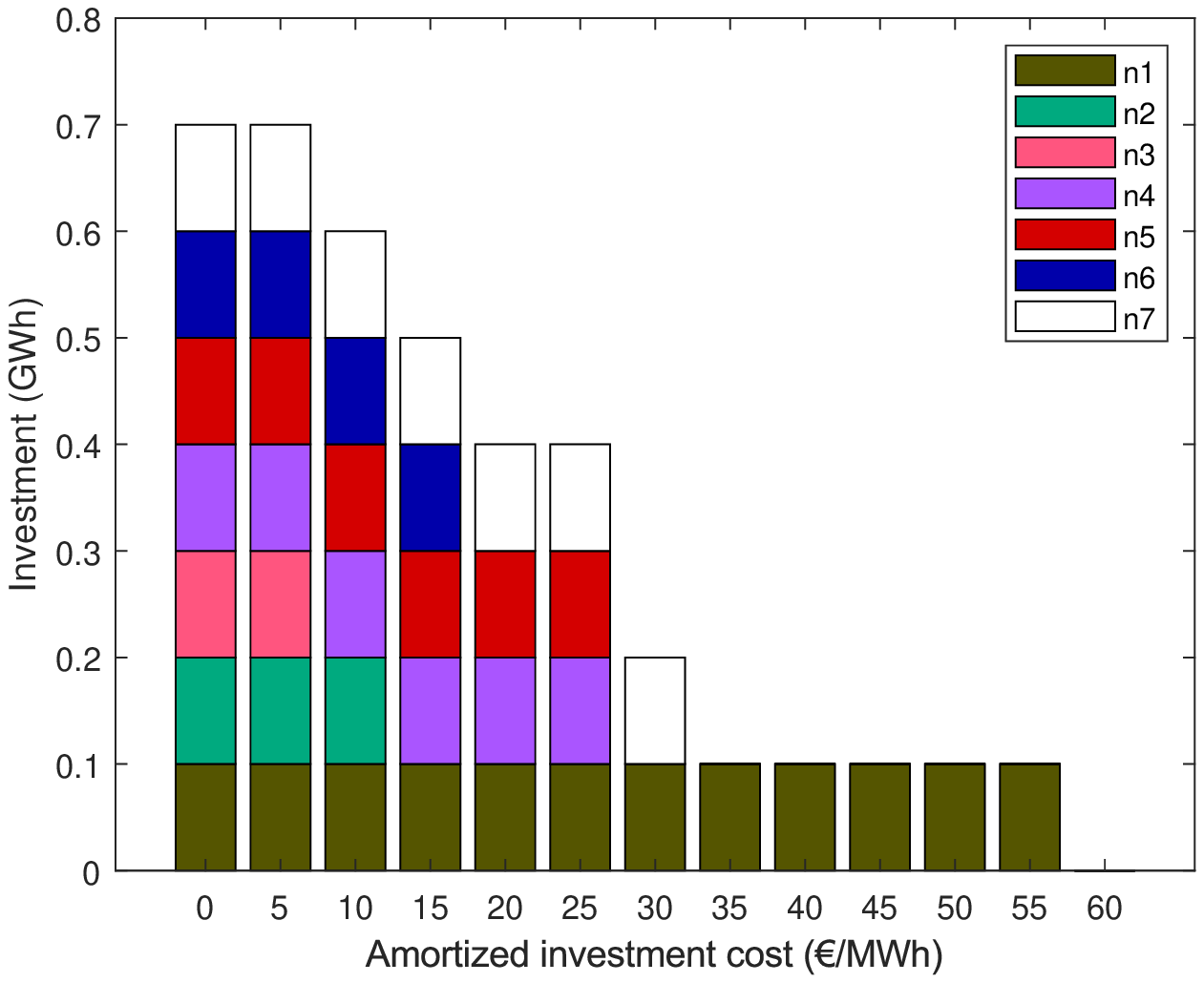}}
\subfigure[Profit-maximizing merchant]{\label{fig:me_co_inv}\includegraphics[width=80mm]{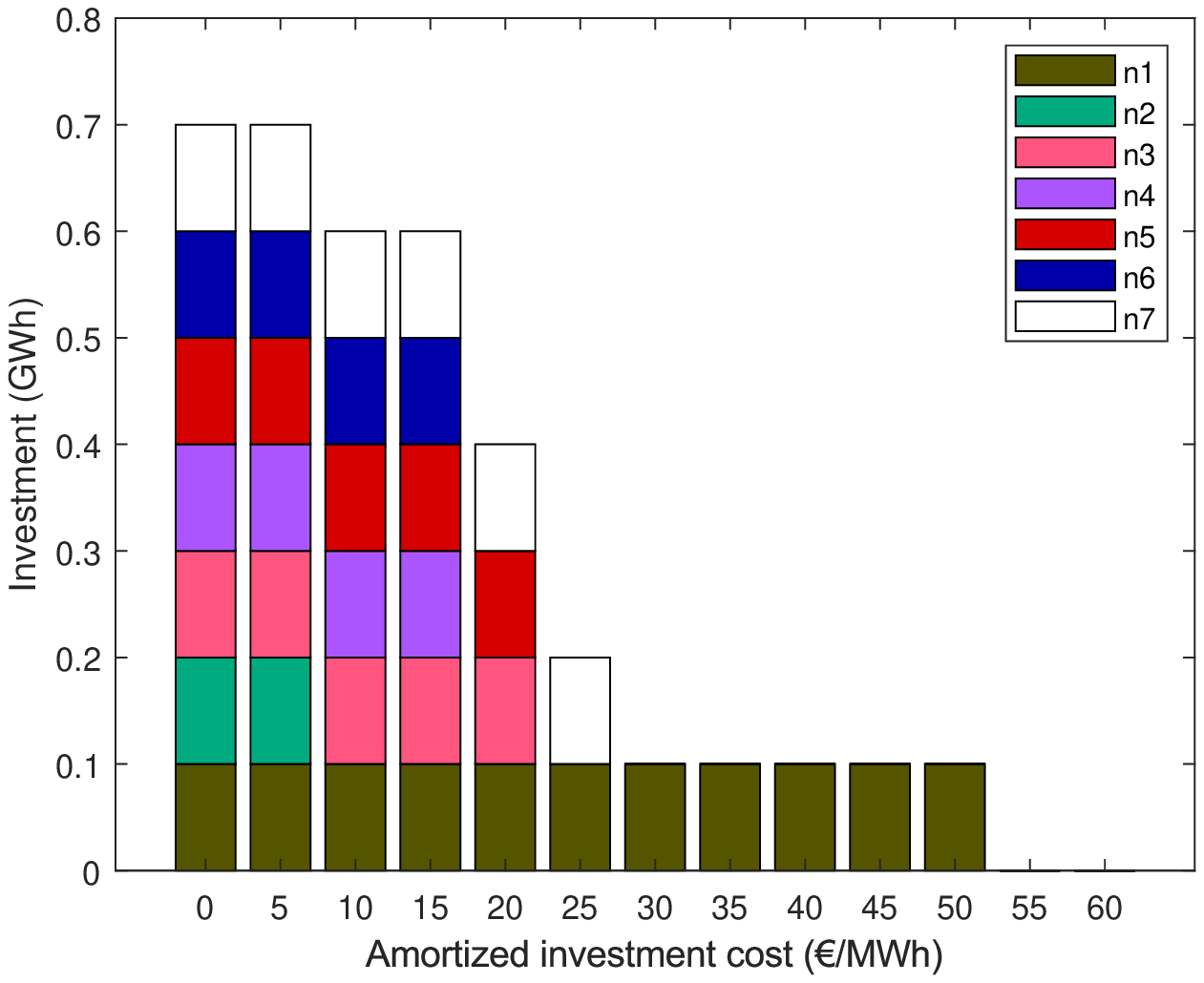}}
\caption{Optimal storage investment sizes and locations under Cournot oligopoly}
\label{fig:co_investments}
\end{figure}

When the lower-level market behaves as a Cournot oligopoly (Fig. \ref{fig:co_investments}), the amortized investment costs need to be lower than in perfect competition. Although market power drives prices higher, the equilibrium prices in this study setting become flatter over time, which decreases any temporal arbitrage opportunities. The optimal investment locations are also reversed: the primary one is $n1$ (Germany), followed by the Dutch nodes. These have lower equilibrium prices and more temporal variation than the others.
Nevertheless, in this setting there are more prominent differences based on the storage investor type. The investments of the welfare-maximizer (Fig. \ref{fig:sw_co_inv}) are greater than or equal to the investments of the merchant (Fig. \ref{fig:me_co_inv}) at the higher end of the cost spectrum. However, \euro15/MWh provides a counterexample with the merchant's total investments of 600 MWh exceeding the total investments of the welfare-maximizer of 500 MWh. The merchant's investments at that specific cost also lead to less power plant ramping but cause more congestion than the welfare-maximizer's investment decision.

The welfare effects of storage under PC are not very sensitive to the investment cost assumption (Table \ref{table:welfare_effects_pc}). As the investment sizes in this battery-based example are considerably small compared to the whole market, the SW gain is relatively minor, but it amplifies with a decreasing investment cost (and increasing total investment size). When the welfare-maximizer and merchant do not invest symmetrically (e.g., \euro65/MWh or \euro35/MWh), the producers are better off when the investor is a merchant. This is because the merchant invests less and, therefore, effects such as price-smoothing do not impact the position of producers as much as a higher capacity would. Storage investments decrease GR (value of imports) except with a merchant investor at \euro65/MWh.

\begin{table}[!h]
\scriptsize
\caption{Welfare effects of storage investments with a selected set of investment costs (\euro/MWh) on social welfare (SW), investor surplus (IS), producer surplus (PS), consumer surplus (CS), and grid revenue (GR) (value of imports) when the lower-level market is perfect competition (PC). Results denote absolute changes to a market without any investments (+/-). }
\centering
\begin{tabular}{c|c|l|l|l|l|l|l}
\label{table:welfare_effects_pc}
     Amortized investment & Model & SW (k\euro) & IS (k\euro) & PS (k\euro) & CS (k\euro) & GR (k\euro) & Investment, \\
     cost (\euro/MWh) &&&&&&& (GWh) \\
         \hline
         \hline
     \textit{no investment} &  PC & 2 201 628.15 & - &  438 683.35
 & 1 749 290.80 & 13 654.01  & -   \\
         \hline
         \hline
     65 &  1. CP / 2. SW-PC & +2.542 & +1.95 & -86.53 & +88.02 & -0.90 & +0.3 \\
        &  3. M-PC   & +2.541 & +2.23 & -82.50 & +81.90 & +0.92 & +0.2 \\  
        \hline
     50 &  1. CP / 2. SW-PC & +7.04 & +6.45 & -86.53 & +88.02 & -0.90 & +0.3 \\
        &  3. M-PC   & +7.04 & +6.45 & -86.53 & +88.02 & -0.90 & +0.3  \\ 
       \hline
     35 &  1. CP / 2. SW-PC & +13.06 & +11.95 & -101.63 & +103.54 & -0.80 & +0.6 \\
        &  3. M-PC   & +12.83 & +11.98 & -90.93  & +93.13  & -1.34 & +0.5  \\ 
\end{tabular}
\end{table}

Again, Cournot oligopoly causes more variation in the results (Table \ref{table:welfare_effects_co}). First, a welfare-maximizing entity may invest in storage (at \euro55/MWh), although the investment causes a loss in terms of IS, because its benefit for producers and consumers exceeds the investment loss and the deficit for GR. Naturally, this would never happen for the merchant investor. Second, at the low end of the investment cost spectrum (e.g., \euro25/MWh), the producers' disadvantage or benefit depends on the investor type. The consumers, although always benefiting from storage, are better off with a merchant investor, unlike the producers. Gains from imports can also increase with a welfare-maximizer at the lower end of costs.

\begin{table}[!h]
\scriptsize
\caption{Welfare effects of storage investments with a selected set of investment costs (\euro/MWh) on social welfare (SW), investor surplus (IS), producer surplus (PS), consumer surplus (CS), and grid revenue (GR) (value of imports) when the lower-level market is Cournot oligopoly (CO). Results denote absolute changes to a market without any investments (+/-).}
\centering
\begin{tabular}{c|c|l|l|l|l|l|l}
\label{table:welfare_effects_co}
     Amortized investment & Model & SW (k\euro) & IS (k\euro) & PS (k\euro) & CS (k\euro) & GR (k\euro) & Investment, \\
     cost (\euro/MWh) &&&&&&& (GWh) \\
         \hline
         \hline
     \textit{no investment} &  CO & 1 923 769.99 & - &  989 457.19   
 & 917 432.27 &  16 880.53  & -   \\
        \hline
        \hline        
     55 &  4. SW-CO         & +0.16 & -0.40 & +0.56 & +0.60 & -0.60 & +0.1  \\
        &  5. M-CO          & - & - & - & - & - & - \\    
        \hline
     50 &  4. SW-CO         & +0.66 & +0.10 & +0.56 & +0.60 & -0.60 & +0.1 \\
        &  5. M-CO          & +0.66 & +0.10 & +0.56 & +0.60 & -0.60 & +0.1  \\           
        \hline      
     25 &  4. SW-CO         & +4.88 & +1.78 & +0.88 & +0.10 & +2.13 & +0.4  \\
        &  5. M-CO          & +3.95 & +2.87 & -0.25 & +1.33 & - & +0.2  \\           
     \hline
     15 &  4. SW-CO         & +9.15 & +6.34 & +0.87 & +0.45 & +1.50 & +0.5  \\
        &  5. M-CO          & +8.43 & +6.92 & -1.05 & +3.27 &- 0.71 & +0.6  \\
\end{tabular}
\end{table}

\paragraph{Transmission Lines}

To study how technical (in)efficiencies impact the results, we analyze one scenario with a 20\% reduction in the transmission capacities and another without transmission limits. Neither the investment sizes nor the locations change under perfect competition with less transmission capacity (Table \ref{table:welfare_effects_network}), although ramping increases and transmission lines are more likely to be congested. Nevertheless, with infinite transmission capacity, both the welfare-maximizer (2. SW-PC) and the merchant (3. M-PC) refrain from making any investments. This is likely because the welfare-maximizer cannot increase the market efficiency from alleviating any congestion. By contrast, the merchant actually benefits from the technical inefficiencies of the system, yielding the highest profit with the reduced transmission capacity. 

\begin{table}[!h]
\scriptsize
\caption{Market equilibrium under perfect competition (PC) or Cournot oligopoly (CO) indicated by social welfare (SW), investor surplus (IS), producer surplus (PS), consumer surplus (CS), and grid revenue (GR) (value of imports). Results denote absolute changes to the base case (+/-). }
\centering
\begin{tabular}{c|c|l|l|l|l|l|l}
\label{table:welfare_effects_network}
     Case & Model & SW (k\euro) & IS (k\euro) & PS (k\euro) & CS (k\euro) & GR (k\euro) & Investment, \\
     &&&&&&& (GWh) \\
         \hline
         \hline
     \textit{Base Case} &  1. CP / 2. SW-PC & 2 201 635.19 & 6.45 &  438 596.82
 & 1 749 378.82 &  13 653.10 & 0.3 \\
     \textit{PC} &  3. M-PC          & 2 201 635.19 & 6.45 &  438 596.82
 & 1 749 378.82 &  13 653.10 & 0.3 \\
       \hline
     \textit{-20\% transmission} &  1. CP / 2. SW-PC & -2 866.77 & +0.12 & -4 350.79   
 & +3 077.34 & -1 593.44 & - \\
     \textit{capacity}  &  3. M-PC            & -2 866.77 & +0.12 & -4 350.79   
 & +3 077.34 & -1 593.44 & - \\
       \hline
     \textit{No transmission} &  1. CP / 2. SW-PC & +19 282.30 & -6.45 &  +115 462.73   
 & -82 520.88 & -13 653.10 & -0.3 \\
     \textit{limits}  &  3. M-PC           & +19 282.30 & -6.45 &  +115 462.73   
 & -82 520.88 & -13 653.10 & -0.3 \\
       \hline
       \hline
      \textit{Base Case} &  4. SW-CO & 1 923 770.65 & 0.10 &  989 457.74   
 & 917 432.87 &  16 879.93 & 0.1 \\
      \textit{CO}  &  5. M-CO & 1 923 770.65 & 0.10 &  989 457.74   
 & 917 432.87 &  16 879.93 & 0.1 \\
       \hline
     \textit{-20\% transmission} &  4. SW-CO & -366.89 & -0.10 &  +4 360.63   
 & -2 456.04 & -2 271.37 & -0.1 \\
     \textit{capacity}  &   5. M-CO   & -367.31 & +0.09 &  +4 363.76   
 & -2 459.26 & -2 271.91 & - \\      
       \hline
     \textit{No transmission} &  4. SW-CO    & -6 737.68 & -0.10 & -42 771.88   
 & +52 914.22 & -16 879.93 & -0.1 \\
     \textit{limits}  &   5. M-CO     & -6 737.68 & -0.10 & -42 771.88   
 & +52 914.22 & -16 879.93 & -0.1 \\
\end{tabular}
\end{table}

Similarly, when the market is imperfect (under Cournot oligopoly), neither of the upper-level decision makers wants to invest with infinite transmission capacity. Intriguingly, with -20\% transmission constraints, the welfare-maximizer (4. SW-CO) does not optimally invest in storage capacity, although the merchant's (5. M-CO) decision remains unchanged from the base case (100 MWh in $n1$). Compared to the base case, the welfare-maximizer's decision slightly increases total ramping and congestion in the system. Nevertheless, the decision not to invest leaves consumer and grid revenue better off than under the merchant, contributing to a higher SW.

\section{Discussion and Conclusions}  \label{discussion}

Earlier research has shown general impacts of having energy storage in the power system. We know that large-scale storage tends to smooth prices and increase social welfare. Technically, it can alleviate network congestion and reduce ramping. In this paper, we have demonstrated that whether storage investments occur in the first place depends both on the investor type and the underlying market setting.

Specifically, a welfare-maximizing storage owner under perfect competition is equivalent to a central-planning standpoint, and it uses storage to mitigate any economic inefficiencies. Indeed, storage investments are the highest for a welfare-maximizer under perfect competition. The merchant's investments are similar but require lower investment costs. Consumers benefit more than the other market participants of the investments and producers are better off with a merchant investor, who never invests more than a welfare-maximizer under perfect competition. Primary locations for investments are nuclear-dominated France and Belgium. 

The characteristics of the underlying market seem to affect investments even more than the investor type: under Cournot oligopoly, the primary investment location is Germany for both investors. Additionally, the investments require much lower investment costs, because temporal arbitrage becomes harder. A welfare-maximizer in most situations invests at least as much as a merchant, although the size varies. Indeed, the welfare-maximizer may invest even if storage itself is not profitable, because it enhances other aspects of the markets enough to increase social welfare. Overall, the results are somewhat conflicting with \cite{Siddiqui2018}, reflecting the complexity that arises from the spatial and temporal aspects in the studied system. 

Nevertheless, because our estimates of investment costs are relatively low, we conclude that the profitability of storage investments is still disputable. In particular, the model considers only day-ahead markets for power production and storage use. In reality, revenue streams from providing ancillary services at reserve markets are likely to be important, too, and should be considered in future work (similar to, e.g., \citet{Xu2017}). To circumvent this limitation, we use low-range estimates for the storage investment cost, although they may still be too optimistic under the current market situation (cf. \citet{IRENA2017}). Additionally, our test network is stylized. Comparing to, e.g., \citet{Xu2017} and \citet{Dvorkin2018} who observe that only a few nodes out of 240 are actually of interest for a storage investment, our network may not capture fine differences which relate to, e.g., high wind capacity (and possibility for negative prices) or local, often congested transmission lines. Nevertheless, our work extends the earlier literature by providing insights into imperfectly competitive markets, as well as the Western European market.

Methodologically, the most important step for future research will be to work on decomposition methods for solving the original MIQCQP model. This would make it possible to incorporate a greater range of investment decisions for the upper-level decision-makers in a realistic study setting. An alternative pathway would be to simplify the underlying market from a QP into a linear program (LP) as cost minimization, such that the resulting bi-level MPPDC would be an MILP. This would make it possible to study larger problems via our chosen methods, but would essentially exclude the investigation of market power within the system. Other possible solution methods could include, e.g., SOS1 (Special Ordered Sets of Type 1) constraints for lower-level complementarity conditions, or parametric approaches for handling bilinear terms, such as \cite{Bylling2019}. Furthermore, modelling uncertainty in hourly VRE generation explicitly, e.g., via stochastic programming, would be another option for future research.

\section*{Acknowledgement}

This paper has benefited from a presentation at the 2018 EURO Conference. We wish to thank Lina Reichenberg of Chalmers University of Technology and Aalto University for providing the code for the clustering model, as well as our colleagues in the Systems Analysis Laboratory at Aalto University for general discussions on the topics of this paper.

\bibliography{main}

\begin{appendices}

\section{Appendix}
\label{appendix_a}

The nonlinear terms in the strong-duality equality Eq. \eqref{SDE_CO} are linearized by replacing them with new positive variables ${x}^\text{lb}_{m,t,n,j,y}$, ${x}^\text{ub}_{m,t,n,j,y}$, ${x}^\text{out}_{m,t,n,j,y}$, and ${x}^\text{in}_{m,t,n,j,y}$. Now, Eq. \eqref{SDE_CO} under Cournot oligopoly\footnote{Under perfect competition, Eq. \eqref{SDE_CO_2} lacks the terms with bold font.} is rewritten as 
\begin{align}
& \sum_{m \in \mathcal{M}} \sum_{t \in \mathcal{T}} \sum_{n \in \mathcal{N}} W_{m}   \, \Bigg[  
\bigg( D^{\text{int}}_{m,t,n} q_{m,t,n} - \frac{1}{2} D_{m,t,n}^{\text{slp}} q_{m,t,n}^2 \bigg) 
\nonumber \\
& \mathbf{ - \frac{1}{2} D_{m,t,n}^{\textbf{slp}} \sum_{i'\in \mathcal{I'}} \bigg( \sum_{u \in \mathcal{U}_{n,i'}} g^\textbf{conv}_{m,t,n,i',u}  + \sum_{\substack{e \in \mathcal{E}}}  g^e_{m,t,n,i'} + r^\textbf{out}_{m,t,n,i'}  - r^\textbf{in}_{m,t,n,i'} \bigg)^2 }
\nonumber \\
&  - \sum_{i'\in \mathcal{I'}} \sum_{u \in \mathcal{U}_{n,i'}}  C_{u}^\text{conv}  g^\text{conv}_{m,t,n,i',u}   -  \sum_{i\in \mathcal{I}} C^{\text{sto}} r^{\text{out}}_{m,t,n,i} \Bigg]
\nonumber \\
\ge 
& \sum_{m \in \mathcal{M}} \sum_{t \in \mathcal{T}} \sum_{n \in \mathcal{N}} W_{m}  \, \Bigg[  
 \frac{1}{2} D_{m,t,n}^{\text{slp}} \bigg(  q_{m,t,n}^2 + \mathbf{ \sum_{i'\in \mathcal{I'}} \bigg( \sum_{u \in \mathcal{U}_{n,i'}} g^\textbf{conv}_{m,t,n,i',u}  + \sum_{\substack{e \in \mathcal{E}}}  g^e_{m,t,n,i'} }
 \nonumber \\ 
& \mathbf{ + r^\textbf{out}_{m,t,n,i'}  - r^\textbf{in}_{m,t,n,i'} \bigg)^2 } \bigg)
 \Bigg] 
\nonumber \\
& + \sum_{m \in \mathcal{M}} \sum_{t \in \mathcal{T}} \sum_{\ell \in \mathcal{L}} T_{t} K_{\ell}  \big( \underline{\mu}_{m,t,\ell} + \overline{\mu}_{m,t,\ell} \big) - \sum_{m \in \mathcal{M}} \sum_{t \in \mathcal{T}} \sum_{n \in \mathcal{N}} \underline{R}_{n,j} \sum_{y \in \mathcal{Y}} {x}^\text{lb}_{m,t,n,j,y}
\nonumber \\
& + \sum_{m \in \mathcal{M}} \sum_{t \in \mathcal{T}} \sum_{n \in \mathcal{N}} \sum_{y \in \mathcal{Y}} {x}^\text{ub}_{m,t,n,j,y}
+ \sum_{m \in \mathcal{M}} \sum_{t \in \mathcal{T}} \sum_{n \in \mathcal{N}} T_{t} R^\text{out}_j \sum_{y \in \mathcal{Y}} {x}^\text{out}_{m,t,n,j,y}
\nonumber \\
&
+ \sum_{m \in \mathcal{M}} \sum_{t \in \mathcal{T}} \sum_{n \in \mathcal{N}}
T_{t} R^\text{in}_j \sum_{y \in \mathcal{Y}} {x}^\text{in}_{m,t,n,j,y}
- \sum_{m \in \mathcal{M}} \sum_{t \in \mathcal{T}} \sum_{n \in \mathcal{N}} \sum_{i'\in \mathcal{I'}} \underline{R}_{n,i'} \overline{R}_{n,i'} {\lambda}^\text{lb,p}_{m,t,n,i'}
\nonumber \\
&
+ \sum_{m \in \mathcal{M}} \sum_{t \in \mathcal{T}} \sum_{n \in \mathcal{N}}
\sum_{i'\in \mathcal{I'}} \overline{R}_{n,i'}{\lambda}^\text{ub,p}_{m,t,n,i'}
+ \sum_{m \in \mathcal{M}} \sum_{t \in \mathcal{T}} \sum_{n \in \mathcal{N}} \sum_{i'\in \mathcal{I'}} T_{t} \,  R^\text{out}_{i'} \,  \overline{R}_{n,i'} {\lambda}^\text{out,p}_{m,t,n,i'}
\nonumber \\
&
+ \sum_{m \in \mathcal{M}} \sum_{t \in \mathcal{T}} \sum_{n \in \mathcal{N}}
\sum_{i'\in \mathcal{I'}}  T_{t} \,  R^\text{in}_{i'} \,  \overline{R}_{n,i'}
{\lambda}^\text{in,p}_{m,t,n,i'}+ \sum_{m \in \mathcal{M}} \sum_{t \in \mathcal{T}} \sum_{n \in \mathcal{N}} \sum_{i'\in \mathcal{I'}} \sum_{\substack{e \in \mathcal{E}}} T_{t} \,  A^e_{m,t,n} \, \overline{G}{}^e_{n,i'} {\beta}^{e}_{m,t,n,i'}
\nonumber \\
&
+ \sum_{m \in \mathcal{M}} \sum_{t \in \mathcal{T}} \sum_{n \in \mathcal{N}} \sum_{i'\in \mathcal{I'}} \sum_{\substack{u \in \mathcal{U}_{n,i'}}}
T_{t} \, R^{\text{down}}_u \, \overline{G}^\text{conv}_{n,i',u}{\beta}^{\text{down}}_{m,t,n,i',u} \nonumber \\
&+ \sum_{m \in \mathcal{M}} \sum_{t \in \mathcal{T}} \sum_{n \in \mathcal{N}}
\sum_{i'\in \mathcal{I'}} \sum_{\substack{u \in \mathcal{U}_{n,i'}}} T_{t} \, R^{\text{up}}_u \, \overline{G}^\text{conv}_{n,i',u} {\beta}^{\text{up}}_{m,t,n,i',u}
\nonumber \\
&
+ \sum_{m \in \mathcal{M}} \sum_{t \in \mathcal{T}} \sum_{n \in \mathcal{N}} \sum_{i'\in \mathcal{I'}} \sum_{\substack{u \in \mathcal{U}_{n,i'}}} T_{t}  \, \overline{G}^\text{conv}_{n,i',u} \beta^\text{conv}_{m,t,n,i',u} \label{SDE_CO_2}
\end{align}

Following the procedure of \citet{Baringo2012}, five new linear constraints for the two additional auxiliary variable are needed to linearize the original bilinear terms. For the term ${\lambda}^\text{lb,m}_{m,t,n,j} \sum_{y \in \mathcal{Y}} z_{n,j,y} \overline{R}^d_{y}$:
\begin{align}
        {x}^\text{lb}_{m,t,n,j,y} &= {\lambda}^\text{lb,m}_{m,t,n,j} \overline{R}^d_{y} - \hat{x}^\text{lb}_{m,t,n,j,y} \label{linearization_first} \\
        z_{n,j,y} \underline{{\lambda}}^\text{lb,m}_{m,t,n,j} \overline{R}^d_{y} &\leq {x}^\text{lb}_{m,t,n,j,y} \\
        {x}^\text{lb}_{m,t,n,j,y} &\leq z_{n,j,y} \overline{{\lambda}}^\text{lb,m}_{m,t,n,j} \overline{R}^d_{y} \\
        (1 - z_{n,j,y}) \underline{{\lambda}}^\text{lb,m}_{m,t,n,j} \overline{R}^d_{y} &\leq \hat{x}^\text{lb}_{m,t,n,j,y} \\
         \hat{x}^\text{lb}_{m,t,n,j,y} &\leq (1 - z_{n,j,y}) \overline{{\lambda}}^\text{lb,m}_{m,t,n,j} \overline{R}^d_{y} 
\end{align}

\noindent where $\hat{x}^\text{lb}_{m,t,n,j,y}$ is an auxiliary variable, $\underline{{\lambda}}^\text{lb,m}_{m,t,n,j}$ denotes the lower and $\overline{{\lambda}}^\text{lb,m}_{m,t,n,j}$ the upper bound for the corresponding dual variable. The constraints ensure that if $z_{n,j,y}=1$ (storage investment is made), the corresponding auxiliary variable $\hat{x}^\text{lb}_{m,t,n,j,y} = 0$ and thus ${x}^\text{lb}_{m,t,n,j,y}$ will equal ${\lambda}^\text{lb,m}_{m,t,n,j} \overline{R}^d_{y}$. On the other hand, if $z_{n,j,y}=0$ (storage investment is not selected), then ${x}^\text{lb}_{m,t,n,j,y}=0$ and it holds that $\underline{{\lambda}}^\text{lb,m}_{m,t,n,j} \overline{R}^d_{y} \le \hat{x}^\text{lb}_{m,t,n,j,y} \le \overline{{\lambda}}^\text{lb,m}_{m,t,n,j} \overline{R}^d_{y}$.

Similarly, for the other bilinear terms, we have
\begin{align}
        {x}^\text{ub}_{m,t,n,j,y} &= {\lambda}^\text{ub,m}_{m,t,n,j} \overline{R}^d_{y} - \hat{x}^\text{ub}_{m,t,n,j,y} \\
        z_{n,j,y} \underline{{\lambda}}^\text{ub,m}_{m,t,n,j} \overline{R}^d_{y} &\leq {x}^\text{ub}_{m,t,n,j,y} \\
        {x}^\text{ub}_{m,t,n,j,y} &\leq z_{n,j,y} \overline{{\lambda}}^\text{ub,m}_{m,t,n,j} \overline{R}^d_{y} \\
        (1 - z_{n,j,y}) \underline{{\lambda}}^\text{ub,m}_{m,t,n,j} \overline{R}^d_{y} &\leq \hat{x}^\text{ub}_{m,t,n,j,y} \\
         \hat{x}^\text{ub}_{m,t,n,j,y} &\leq (1 - z_{n,j,y}) \overline{{\lambda}}^\text{ub,m}_{m,t,n,j} \overline{R}^d_{y} 
\end{align}
\begin{align}
        {x}^\text{out}_{m,t,n,j,y} &= {\lambda}^\text{out,m}_{m,t,n,j} \overline{R}^d_{y} - \hat{x}^\text{out}_{m,t,n,j,y} \\
        z_{n,j,y} \underline{{\lambda}}^\text{out,m}_{m,t,n,j} \overline{R}^d_{y} &\leq {x}^\text{out}_{m,t,n,j,y} \\
        {x}^\text{out}_{m,t,n,j,y} &\leq z_{n,j,y} \overline{{\lambda}}^\text{out,m}_{m,t,n,j} \overline{R}^d_{y} \\
        (1 - z_{n,j,y}) \underline{{\lambda}}^\text{out,m}_{m,t,n,j} \overline{R}^d_{y} &\leq \hat{x}^\text{out}_{m,t,n,j,y} \\
         \hat{x}^\text{out}_{m,t,n,j,y} &\leq (1 - z_{n,j,y}) \overline{{\lambda}}^\text{out,m}_{m,t,n,j} \overline{R}^d_{y} 
\end{align}
\begin{align}
        {x}^\text{in}_{m,t,n,j,y} &= {\lambda}^\text{in,m}_{m,t,n,j} \overline{R}^d_{y} - \hat{x}^\text{in}_{m,t,n,j,y} \\
        z_{n,j,y} \underline{{\lambda}}^\text{in,m}_{m,t,n,j} \overline{R}^d_{y} &\leq {x}^\text{in}_{m,t,n,j,y} \\
        {x}^\text{in}_{m,t,n,j,y} &\leq z_{n,j,y} \overline{{\lambda}}^\text{in,m}_{m,t,n,j} \overline{R}^d_{y} \\
        (1 - z_{n,j,y}) \underline{{\lambda}}^\text{in,m}_{m,t,n,j} \overline{R}^d_{y} &\leq \hat{x}^\text{in}_{m,t,n,j,y} \\
         \hat{x}^\text{in}_{m,t,n,j,y} &\leq (1 - z_{n,j,y}) \overline{{\lambda}}^\text{in,m}_{m,t,n,j} \overline{R}^d_{y} \label{linearization_last}.
\end{align}

\end{appendices}

\end{document}